\documentclass[12pt]{article}
\usepackage{latexsym}
\usepackage{amsmath}
\usepackage{epsfig,graphics}

\newcommand{\be}{\begin{equation}}
\newcommand{\ee}{\end{equation}}

\newlength{\figsize}
\begin{document}

\begin{titlepage}

\vspace*{0.7in}
 
\begin{center}
{\large\bf Closed k-strings in SU(N) gauge theories : 2+1 dimensions \\}
\vspace*{1.0in}
{Barak Bringoltz$^{a}$ and Michael Teper$^{b}$\\
\vspace*{.2in}
$^{a}$Department of Physics, University of Washington, Seattle,
WA 98195-1560, USA\\
\vspace*{.1in}
$^{b}$ Rudolf Peierls Centre for Theoretical Physics, University of Oxford,\\
1 Keble Road, Oxford OX1 3NP, UK\\
}
\end{center}

\vspace*{0.55in}

\begin{center}
{\bf Abstract}
\end{center}

We calculate the ground state energies of closed $k$-strings 
in (2+1)-dimensional SU(N) gauge theories, for $N=4,5,6,8$
and $k=2,3,4$. From the dependence of the ground state energy 
on the string length, we infer that such $k$-strings are
described by an effective string theory that is in the same 
bosonic universality class (Nambu-Goto) as the fundamental
string. When we compare the continuum $k$-string tensions
to the corresponding fundamental string tensions, we find
that the ratios are close to, but typically $1\%-2\%$ above,
the Casimir scaling values favoured by some theoretical
approaches. Fitting the $N$-dependence in a model-independent
way favours an expansion in $1/N$ (as in Casimir scaling)
rather than the $1/N^2$ that is suggested by naive colour 
counting. We also observe that the low-lying spectrum of $k$-string
states falls into sectors that belong to particular irreducible
representations of $SU(N)$, demonstrating that the 
dynamics of string binding knows about the full gauge group and 
not just about its centre.

\end{titlepage}

\setcounter{page}{1}
\newpage
\pagestyle{plain}

\section{Introduction}
\label{section_intro}

In this paper we calculate the ground state energy of a confining 
flux loop that winds once around a spatial torus of
length $l$, with the flux in a higher representation
than the fundamental. We do so for a number of SU($N$) groups
in $D=2+1$ dimensions. This study complements recent
calculations for fundamental flux loops
\cite{gs_k1}
and the excited state spectra of these
\cite{es_k1}.
In a forthcoming paper we shall analyse the excited state spectrum 
of multiply wound flux loops 
\cite{es_k}
and similar calculations in $D=3+1$ dimensions are under way.

These numerical calculations have a number of motivations.
Firstly there are old ideas that the relevant degrees of
freedom of linearly confining SU($N$) gauge theories are
string-like flux tubes and that these should be described by an
effective string theory
\cite{polyakov}
which might describe all the important physics of the field theory, 
particularly as $N\to\infty$
\cite{thooft}.
Calculations such as ours can serve to test this programme;
and, at the very least, to pin down the details of the effective
string theory that describes the dynamics of long flux tubes
\cite{LSW,PS}.
Secondly, there has been dramatic recent progress in the 
construction of string duals to various supersymmetric field 
theories
\cite{maldacena}.
There is a great effort to extend this to gauge theories and to QCD
(see
\cite{zamaklar_evans}
for recent reviews). Our calculations may provide useful 
information in the search for the appropriate construction.

Our recent calculation 
\cite{gs_k1,es_k1}
of the spectrum of a closed loop of fundamental flux as a function 
of its length $l$, demonstrated that this spectrum is accurately
given by the simple Nambu-Goto string model
\cite{arvis}
down to lengths much smaller than those at which an
expansion in powers of $1/\sigma l^2$ diverges. 
While this does not contradict  the conventional approach
of starting with a background consisting of a long `straight' string 
and expanding in fluctuations around it
\cite{LSW}, 
it does imply that one can do considerably better by expanding instead
in small corrections around the full Nambu-Goto solution. Hints 
of this possibility can already be discerned in some recent work, 
\cite{LW,JD},
based on 
\cite{LSW,PS}
respectively, that showed that the next term beyond the Luscher
term
\cite{LSW}
in the expansion in powers of  $1/\sigma l^2$ is also universal
and has the same coefficient as in the Nambu-Goto case.
This fact, that we can identify the states as being essentially
those of the Nambu-Goto model with
the corresponding `phonon' occupation numbers, means that the
observed small deviations from Nambu-Goto have the potential 
to provide detailed and powerful constraints on the form of the 
additional string interactions that are needed.

The $k$-string that is the object of our study in this paper,
can be thought of as a bound state of $k$ fundamental 
strings. Earlier calculations have shown that 
for $N<\infty$ there do indeed exist such bound states
both in $D=2+1$ and in $D=3+1$
\cite{old_k_oxford,blmtuw-glue04,old_k_pisa,ckahmN}.
This binding cannot be readily incorporated into the usual 
analytic frameworks
\cite{LSW,PS}
since the binding interaction presumably involves the exchange of 
small closed loops of string that are not under analytic control. 
Thus such strings are of particular interest.

In the case of SU($N$) gauge theories in $D=2+1$ there is an
additional motivation. There has been significant recent progress 
towards an analytic solution of these theories using 
a variational Hamiltonian approach
\cite{nair}.
(See also related work in
\cite{nair_other}.) These calculations make predictions for both
the fundamental string tension and for $k$-string tensions.
In 
\cite{gs_k1}
we found that the former prediction is very accurate and improves
with increasing $N$, being only about $1\%$ off the correct value
at $N=\infty$. (Although this discrepancy is statistically
very significant.) In this paper we shall provide a similar test
for the predicted  $k$-string tensions. 

In the next section we describe our method for calculating 
the energy of a closed $k$-string. We then present, in
Section~\ref{section_charge}, our most accurate results 
for the $l$-dependence of this energy (which happens to be for
the case $k=2$) and compare to the expectation of the simple 
bosonic string model. In Section~\ref{section_continuum}
we calculate the $k$-string tensions, extrapolate these
to the continuum limit, and test the conjecture that they
scale with the appropriate Casimir. We follow this,
in Section~\ref{section_N}, with a model-independent
analysis of the $N$-dependence of these string tensions.
Finally, in Section~\ref{section_group}, we look at the
wave-functions of the ground and excited states and
show that in fact they fall into representations of the
group, rather than being determined simply by the center.
A summary of this work has been presented in
\cite{lat07}.

\section{Methodology}
\label{section_methods}

Consider a spatial torus of length $l$ with all the other tori 
so large that they may be considered infinite. If the system
is linearly confining, any flux around the torus will be confined
to `string-like' flux tubes. If the source from which such a flux
would emanate transforms as $\psi \to z^k \psi$ under a gauge
transformation, $z$, that belongs to the center, $Z_N$, of the SU($N$) 
gauge group, the flux tube is called a $k$-string. This is a useful
categorisation since it is invariant under screening by gluons.
For large $l$ the ground state $k$-string energy $E_{0,k}(l)$ gives 
us the $k$-string tension, 
\begin{equation}
\sigma_k = \lim_{l\to\infty} \frac{E_{0,k}(l)}{l} .
\label{eqn_sigmak} 
\end{equation} 
(In this paper $\sigma_k$ is the tension of the ground state
$k$-string, $\sigma_f \equiv \sigma_{k=1}$ is the fundamental
string tension,$\sigma_{\cal{R}}$ is the tension of a string
carrying flux in representation $\cal{R}$, and $\sigma$ is
a generic string tension. When there is no ambiguity, we drop
the extra $k$ label on $E_0$.)
Whether stable $k$-strings know about the full group or only about 
the center of the group, is one of the issues we address in this
paper. 

As one varies $l$ there will be a finite volume phase transition at
$l_c=1/T_c$, where  $T_c$ is the value of the deconfining temperature. 
For $l \geq l_c$ the flux around the torus will be in a flux tube of 
length $l$ and so we can ask what is the effective string theory that 
describes its spectrum. For $l < l_c$ there is no winding flux tube 
about which to ask such a question.

The alternative approach of studying open strings using Wilson loops
(or pairs of unsmeared Polyakov loops),
is complicated by the contribution of the sources to the corresponding
ground state energy. The closed string setup provides a 
theoretically cleaner framework for studying the effective string
theory although for smaller $N$, where the phase transition is second 
order or weakly first order, the influence of the nearby critical
point needs to be taken into account once $l \sim l_c$. On the other
hand open strings ending on point sources provide a way to address
the important question of how perturbative physics at short distances
evolves into non-perturbative physics at large distances, and whether
there is a non-analyticity in this evolution as $N\to\infty$.
Comparing the SU(5) potential energy calculated in
\cite{meyerV}
with the lower $N$ potentials calculated in
\cite{majumdarV}
points to the possibility of an interesting answer to this question.

Possible effective string theories fall into universality classes. 
If we expand $E_0(l)$ in powers of $1/\sigma l^2$ around $\sigma l$,
then for a bosonic string, where the only zero modes are those that 
arise from the $D-2=1$ transverse translations, the first two 
correction terms are universal 
\cite{LW,JD}
\begin{equation}
E_0(l) 
=
\sigma l - \frac{c\pi}{6l} 
- \frac{c^2\pi^2}{72\sigma l^3} + \ldots
\qquad ; \quad c=1 .
\label{eqn_univ} 
\end{equation}
More generally, if there are additional zero modes along the string, 
then we may have $c \neq 1$ (although in that case only the first 
correction term has been shown to be universal). The simplest example 
of a bosonic string
is the Nambu-Goto free string model, whose ground state
energy is
\cite{arvis}
\begin{equation}
E_0(l) 
=
\sigma l 
\left( 
1 - \frac{\pi}{3} \frac{1}{\sigma l^2}
\right)^\frac{1}{2}
\label{eqn_NG} 
\end{equation}
which reproduces eqn(\ref{eqn_univ}) when expanded in powers
of $1/\sigma l^2$.

Our calculations are performed using standard lattice tachniques. 
We work on periodic hypercubic $L_x\times L_y\times L_t$ 
lattices with lattice spacing $a$. The degrees of freedom are 
SU($N$) matrices, $U_l$, assigned to the links $l$ of the lattice.
The partition function is 
\begin{equation}
{\cal Z}(\beta)
=
\int \prod_l dU_l 
e^{-\beta \sum_p\{ 1 - \frac{1}{N}{\mathrm{ReTr}}U_p\}}
\ \ \ \ ; \ \ \ \
\beta=\frac{2N}{a g^2}
\label{eqn_lattice} 
\end{equation} 
where $U_p$ is the ordered product of matrices around the
boundary of the elementary square (plaquette) labelled by $p$
and $g^2$ is the coupling, which in $D=2+1$ has dimensions
of mass. This is the standard Wilson plaquette action and 
the continuum limit is 
approached by tuning $\beta = 2N/ag^2 \to \infty$. 
One expects that for large $N$ physical masses
will be proportional to the 't Hooft coupling 
$\lambda \equiv g^2 N$, and this is indeed what one observes
\cite{glued3}.
So if we vary  $\beta \propto N^2$ then we keep
the lattice spacing $a$  fixed in physical units.

We will consider a loop of flux that winds once around the 
$x$-torus, so that it is of length $l=aL_x$. A generic operator
$\phi_l$ that couples to such a periodic flux loop is an 
ordered product of link matrices along a space-like curve 
that winds once around the $x$-torus, with the link
matrices in the same representation as the flux. 
Correlations are taken in the $t$ direction so that
the energy of the loop is an eigenstate of the Hamiltonian
(transfer matrix) defined on the $xy$ space. 
Such a correlator may be expanded 
\begin{equation}
C(t) 
\equiv \langle \phi^\dagger_l(t) \phi(0) \rangle
=
\sum_{n=0}  |\langle n |\phi(0) | vac \rangle|^2 e^{-E_n t}
\ \ \ \ ; \ \ \ \
E_i \leq E_{i+1}
\label{eqn_corrln} 
\end{equation} 
where, if we have confinement, $E_0$ is the energy of the 
lightest flux loop. Since the fluctuations that determine the 
error in the Monte Carlo calculation of $C(t)$ may be expressed
as a correlation function with a disconnected piece, the
error is approximately independent of $t$ while $C(t)$ itself
decreases exponentially with $t$. Thus one needs operators
$\phi_l$ that have a large projection onto the
desired state, so that this state dominates $C(t)$ at small $t$
where the statistical noise is still relatively small. 
This can be achieved using standard smearing/blocking/variational 
techniques (see e.g. 
\cite{glued3,blmtuw-glue04}).
If we denote a Polyakov loop in the fundamental $k=1$ representation 
by $l_p$, then typical operators for $k$-strings will be  
$\phi_l = \mathrm{Tr}\{l_p^{k-j}\} \mathrm{Tr}\{l_p\}^j$
and we include all such operators in our variational basis,
including their smeared/blocked versions. 

Since $E_0(l)$ becomes large for large $l$,
it can only be accurately calculated for modest values of
$l\surd\sigma$ and so we need to know what are the important finite
$l$ corrections to the linear term, $E_0(l) \simeq \sigma l$
if we are to extract a reliable value for $\sigma$.
This is even more important for heavier strings with $k>1$,
since these are heavier than the fundamental $k=1$ string. 
In 
\cite{gs_k1}
it was shown that for the fundamental string, corrections to
eqn(\ref{eqn_NG}) are extremely small, even very close to $l_c$
and one can safely extrapolate using eqn(\ref{eqn_NG})
from, say, $l\surd\sigma_f \sim 3$.
However there is no guarantee that this will continue to be
the case for $k>1$. Indeed we expect that at fixed $k$ the
binding will vanish as $N\to\infty$, and so in that limit,
at any fixed $l$ however large, the lightest $k$-string will actually
be $k$ fundamental strings, each satisfying eqn(\ref{eqn_univ})
\cite{hmmt}.
Moreover, the fact that all the strings must share the same critical
length, $l_c$, leads us to expect much larger  deviations from
eqn(\ref{eqn_NG}) at small $l$ for $k>1$. In addition there have
been suggestions 
\cite{gliozzi}
that $k>1$ strings do not belong to the bosonic string
universality class and have $c>1$. Thus it is important
to determine the $l$-dependence of the ground state $k$-string
and this is the issue to which we now turn.

\section{Central charge and $l$ dependence of $k$-strings}
\label{section_charge}

We have performed finite volume studies for $N=2,3,4,5,6,8$
at various values of the lattice spacing $a$. Here we shall
present two of our most accurate studies for $k=2$ strings. 
These are for SU(4) at $\beta=32.0$ and  SU(5) at $\beta=80.0$.
In units of the fundamental string tension, the lattice 
spacings correspond to $a\surd\sigma_f \simeq 0.2153$
and $a\surd\sigma_f \simeq 0.1299$ respectively. 

In Table~\ref{table_V} we present the $k=2$ ground state energy
as a function of $l=aL_x$. We also present, for comparison,
the corresponding energy of the fundamental $k=1$ string.
The critical value of $l$ at which one loses the string is
given by $l_c \sim 5 a $ and  $l_c \sim 8 a $ 
for SU(4) and SU(5) respectively.
Thus our range of closed string lengths extends down to nearly
the minimimum possible value. 

There are many useful ways to analyse the $l$-dependence.
Here we shall fit neighbouring values of $l$ to a
Nambu-Goto formula with an `effective' central charge:
\begin{equation}
E_0(l) 
=
\sigma l 
\left( 
1 - c_{eff} \frac{\pi}{3} \frac{1}{\sigma l^2}
\right)^\frac{1}{2}
\label{eqn_NGceff} 
\end{equation}
This will tell us what is the universality class of the
effective string theory that describes $k=2$ flux tubes;
in particular if $\lim_{l\to\infty} c_{eff} = 1$ then
this universality class is that of the simple bosonic
string, just as it is for fundamental strings. In addition
the rate of approach of $c_{eff}(l)$ to its asymptotic value
will tell us something about the size of the corrections.

In Fig.~\ref{fig_NGceffN4} we plot the value of $c_{eff}$ versus $l$ 
for the SU(4) calculation. We show values for the $k=2$ string
and also, for comparison, for the fundamental $k=1$ string.
We express the length $l$ in units of the string tension  $\sigma_k$.
While this is the appropriate variable to use,  as we see from 
eqn(\ref{eqn_NGceff}), it does mean that the scale in fixed
physical units is slightly different for the $k=1$ and $k=2$ analyses.
The horizontal error bars end at the two values of $l$ used
in the fits to eqn(\ref{eqn_NGceff}).
In  Fig.~\ref{fig_NGceffN5} we have a corresponding plot for SU(5).

We observe in  Figs.\ref{fig_NGceffN4},\ref{fig_NGceffN5}
quite strong evidence for the claim that at long distances
the $k=2$ flux tube behaves like a simple bosonic string, just like
the fundamental flux tube. This contradicts the conjecture in 
\cite{gliozzi}
that $\lim_{l\to\infty} c_{eff} = \sigma_k/\sigma_f$. (This latter 
ratio is, from our fits, about 1.35 and 1.52 for SU(4) and  
SU(5) respectively.)

However it is clear that the corrections to Nambu-Goto are very
much larger for $k=2$ than for $k=1$. Indeed, if we limited
our analysis to $l\surd\sigma_k < 2.5$ we might be led
to agree with the conjecture of  
\cite{gliozzi}.
One way to quantify the size of the deviation is to add to
eqn(\ref{eqn_NG}) a correction term that is of higher order than 
the universal terms in eqn(\ref{eqn_univ}), i.e.
\begin{equation}
E_0(l) 
=
\sigma l 
\left( 
1 - \frac{\pi}{3} \frac{1}{\sigma l^2}
+ c^\prime \frac{1}{(\sqrt{\sigma} l)^5}
\right)^\frac{1}{2} .
\label{eqn_NGcorr} 
\end{equation}
Fitting, for example, our SU(5) loop masses gives 
\begin{equation}
c^\prime = \left\{
\begin{array}
{r@{\qquad : \qquad}l}
-0.09(2) & k=1 \\
-0.93(5) & k=2 
\end{array}
\right.
\label{eqn_cprime} 
\end{equation}
What is interesting about this result is that we see that
it is not the $k=2$ string that has an anomalously large
non-universal correction, but rather it is the $k=1$ string
that has an  anomalously small non-universal correction.
A value $c^\prime = O(1)$ is precisely
what one might expect for a generic effective string model
in the Nambu-Goto universality class and that is what we
find for the $k=2$ flux tube. This observation appears
to be both insensitive to the value of $N$ and to the
size of $a$: that is to say, it applies to the continuum limit.

Although our calculations of $E_0(l)$ for $k > 2$ are not accurate
enough to provide  evidence as compelling as for $k=2$, they are 
consistent with $k > 2$ strings also belonging to the Nambu-Goto
universality class, and so we shall assume that to be the case.
Within our statisitical errors it also appears justified
to assume that the dominant finite-$l$ corrections are those
in eqn(\ref{eqn_NG}).

\section{Continuum string tensions and Casimir scaling}
\label{section_continuum}

The prediction of 
\cite{nair}
is that string tensions in D=2+1 SU($N$) gauge theories  should 
be proportional to the quadratic Casimir of the representation of 
the flux. This is, of course, the exact result in D=1+1, since
in that case linear confinement is the Coulomb interaction.
Based on the idea of effective dimensional reduction driven by
a highly disordered vacuum, it was conjectured a long time ago 
\cite{CSold}
that this might also hold in $D=2+1$ and $D=3+1$. There is
some additional evidence for this hypothesis from calculations 
of the potential between charges in various representations
of SU(3)
\cite{BaliDeldar}.

For a given $k$ the smallest Casimir arises for the totally
antisymmetric representation, and this should therefore provide
the ground state $k$-string tension:
\begin{equation}
\frac{\sigma_k}{\sigma_f} 
=
\frac{k(N-k)}{N-1}.
\label{eqn_CS} 
\end{equation}
This is the part of the `Casimir Scaling' hypothesis that we shall be 
mainly testing
in this paper. For this purpose it is useful to have an alternative
conjecture that possesses the correct general properties. A convenient 
and well-known example is provided by the trigonometric form
\begin{equation}
\frac{\sigma_k}{\sigma_f} 
=
\frac{\sin\frac{k\pi}{N}}{\sin\frac{\pi}{N}} 
\label{eqn_MQCD} 
\end{equation}
that was originally suggested on the basis of an M-theory approach to
QCD
\cite{MQCD}.

In fact the full prediction of  
\cite{nair}
for $\sigma_k$ is more specific than eqn(\ref{eqn_CS}), since it
also predicts a value for $\sigma_f$ in terms of $g^2$, and 
including this gives:
\begin{equation}
\frac{\sigma_k}{(g^2N)^2} 
=
\frac{k(N-k)(N+1)}{N^2} \frac{1}{8\pi}.
\label{eqn_kNair} 
\end{equation}
Since we already know that the prediction for the $k=1$ (fundamental)
string tension is too high by about $1-2\%$ (depending on $N$)
\cite{gs_k1}, 
testing eqn(\ref{eqn_kNair}) and eqn(\ref{eqn_CS}) does not come to 
exactly the same thing.

We now turn to our calculations of $\sigma_k$. We will illustrate
these using the example of SU(5) and then state our results for
other values of $N$.

The first step is to extract ground state energies from the Euclidean
correlation functions that we obtain as a result of the variational
calculation described in Section~\ref{section_methods}.
The standard method is to fit the ground state correlator
with a single exponential, $|c|^2 \exp\{-E_o(l) t\}$, for $t\geq t_0$,
and to find the minimum value of $t_0$ that provides an acceptable
fit. (Usually the upper limit of the range in $t$ is chosen so as to
exclude the very noisy values at large $t$.) The value of the
ground state energy obtained by such single-exponential fits we
label by $S$. Of course, since in general $|c|^2 < 1 $, we know 
that there must be some contribution even at $t\geq t_0$ from the 
excited states in eqn(\ref{eqn_corrln}). In the range of $t$ fitted
this contribution will be masked by the statistical errors, but
will lead to a systematic downward shift in all extracted values
of $E_0(l)$. Ideally this shift should be within the statistical
errors but this is far from guaranteed given our
inevitably imperfect control of the statistical analysis.
To bound this systematic error, we also perform fits with
two exponential terms where the energy of the excited state
is chosen to be that of the first excited state (using $S$ fits)
with the same quantum numbers. This will usually provide an upper 
bound on the excited state correction to the ground state energy.
We label the ground state energy obtained by such double-exponential 
fits by $D$. (Note that this choice will not necessarily bound
the effect of such contributions to the $ratio$ of string tensions.)

We calculate the values of $E_0(l)$ for strings of various $k$
on lattices with spatial tori that satisfy $l\surd\sigma_f > 3$. 
We have seen in Section~\ref{section_charge} that at such
distances the Nambu-Goto expression in  eqn(\ref{eqn_NG}) 
accurately represents the $l$-dependence of the ground state
string energy (within statistical errors that are typical of the
present calculation). We therefore use  eqn(\ref{eqn_NGcorr})
to extract values of $\sigma_k$ from the corresponding $E_0(l)$.

Having obtained values of  $\sigma_k$ at various values of
$\beta$ we now want to extrapolate the dimensionless ratios
$\sigma_k/\sigma_f$ to the continuum limit. We do so using:
\begin{equation}
\frac{\sigma_k(a)}{\sigma_f(a)} 
=
\frac{\sigma_k(0)}{\sigma_f(0)} 
+ c_1 a^2\sigma_f
+ c_2 a^4\sigma^2_f
+ \ldots
\label{eqn_cont} 
\end{equation}
where the $c_i$ can be treated as constants over our range of 
$\beta$. Typically extrapolations of this kind include just the 
$O(a^2)$ correction, and one drops the points at the largest values
of $a$ until a statistically acceptable fit is possible.
We shall also perform a $O(a^4)$ fit so as to estimate the
systematic error in the continuum extrapolation.

\begin	{figure}[p]
\begin	{center}
\leavevmode
\setlength{\unitlength}{0.240900pt}
\ifx\plotpoint\undefined\newsavebox{\plotpoint}\fi
\sbox{\plotpoint}{\rule[-0.200pt]{0.400pt}{0.400pt}}%
\begin{picture}(1800,990)(0,0)
\font\gnuplot=cmr10 at 12pt
\gnuplot
\sbox{\plotpoint}{\rule[-0.200pt]{0.400pt}{0.400pt}}%
\put(375.0,150.0){\rule[-0.200pt]{4.818pt}{0.400pt}}
\put(350,150){\makebox(0,0)[r]{\ \ {$1.4$}}}
\put(1705.0,150.0){\rule[-0.200pt]{4.818pt}{0.400pt}}
\put(375.0,348.0){\rule[-0.200pt]{4.818pt}{0.400pt}}
\put(350,348){\makebox(0,0)[r]{\ \ {$1.45$}}}
\put(1705.0,348.0){\rule[-0.200pt]{4.818pt}{0.400pt}}
\put(375.0,545.0){\rule[-0.200pt]{4.818pt}{0.400pt}}
\put(350,545){\makebox(0,0)[r]{\ \ {$1.5$}}}
\put(1705.0,545.0){\rule[-0.200pt]{4.818pt}{0.400pt}}
\put(375.0,743.0){\rule[-0.200pt]{4.818pt}{0.400pt}}
\put(350,743){\makebox(0,0)[r]{\ \ {$1.55$}}}
\put(1705.0,743.0){\rule[-0.200pt]{4.818pt}{0.400pt}}
\put(375.0,940.0){\rule[-0.200pt]{4.818pt}{0.400pt}}
\put(350,940){\makebox(0,0)[r]{\ \ {$1.6$}}}
\put(1705.0,940.0){\rule[-0.200pt]{4.818pt}{0.400pt}}
\put(375.0,150.0){\rule[-0.200pt]{0.400pt}{4.818pt}}
\put(375,100){\makebox(0,0){\ {$0$}}}
\put(375.0,920.0){\rule[-0.200pt]{0.400pt}{4.818pt}}
\put(586.0,150.0){\rule[-0.200pt]{0.400pt}{4.818pt}}
\put(586,100){\makebox(0,0){\ {$0.025$}}}
\put(586.0,920.0){\rule[-0.200pt]{0.400pt}{4.818pt}}
\put(797.0,150.0){\rule[-0.200pt]{0.400pt}{4.818pt}}
\put(797,100){\makebox(0,0){\ {$0.05$}}}
\put(797.0,920.0){\rule[-0.200pt]{0.400pt}{4.818pt}}
\put(1008.0,150.0){\rule[-0.200pt]{0.400pt}{4.818pt}}
\put(1008,100){\makebox(0,0){\ {$0.075$}}}
\put(1008.0,920.0){\rule[-0.200pt]{0.400pt}{4.818pt}}
\put(1219.0,150.0){\rule[-0.200pt]{0.400pt}{4.818pt}}
\put(1219,100){\makebox(0,0){\ {$0.1$}}}
\put(1219.0,920.0){\rule[-0.200pt]{0.400pt}{4.818pt}}
\put(1430.0,150.0){\rule[-0.200pt]{0.400pt}{4.818pt}}
\put(1430,100){\makebox(0,0){\ {$0.125$}}}
\put(1430.0,920.0){\rule[-0.200pt]{0.400pt}{4.818pt}}
\put(1641.0,150.0){\rule[-0.200pt]{0.400pt}{4.818pt}}
\put(1641,100){\makebox(0,0){\ {$0.15$}}}
\put(1641.0,920.0){\rule[-0.200pt]{0.400pt}{4.818pt}}
\put(375.0,150.0){\rule[-0.200pt]{325.215pt}{0.400pt}}
\put(1725.0,150.0){\rule[-0.200pt]{0.400pt}{190.311pt}}
\put(375.0,940.0){\rule[-0.200pt]{325.215pt}{0.400pt}}
\put(150,845){\makebox(0,0){\Large{${{\sigma_{k=2}}\over{\sigma_f}}$}}}
\put(1025,25){\makebox(0,0){\large{$a^2\sigma_f$}}}
\put(375.0,150.0){\rule[-0.200pt]{0.400pt}{190.311pt}}
\put(1591.0,433.0){\rule[-0.200pt]{0.400pt}{37.580pt}}
\put(1581.0,433.0){\rule[-0.200pt]{4.818pt}{0.400pt}}
\put(1581.0,589.0){\rule[-0.200pt]{4.818pt}{0.400pt}}
\put(1241.0,525.0){\rule[-0.200pt]{0.400pt}{26.740pt}}
\put(1231.0,525.0){\rule[-0.200pt]{4.818pt}{0.400pt}}
\put(1231.0,636.0){\rule[-0.200pt]{4.818pt}{0.400pt}}
\put(1025.0,642.0){\rule[-0.200pt]{0.400pt}{26.258pt}}
\put(1015.0,642.0){\rule[-0.200pt]{4.818pt}{0.400pt}}
\put(1015.0,751.0){\rule[-0.200pt]{4.818pt}{0.400pt}}
\put(777.0,622.0){\rule[-0.200pt]{0.400pt}{14.936pt}}
\put(767.0,622.0){\rule[-0.200pt]{4.818pt}{0.400pt}}
\put(767.0,684.0){\rule[-0.200pt]{4.818pt}{0.400pt}}
\put(608.0,639.0){\rule[-0.200pt]{0.400pt}{11.804pt}}
\put(598.0,639.0){\rule[-0.200pt]{4.818pt}{0.400pt}}
\put(598.0,688.0){\rule[-0.200pt]{4.818pt}{0.400pt}}
\put(517.0,607.0){\rule[-0.200pt]{0.400pt}{8.431pt}}
\put(507.0,607.0){\rule[-0.200pt]{4.818pt}{0.400pt}}
\put(507.0,642.0){\rule[-0.200pt]{4.818pt}{0.400pt}}
\put(463.0,641.0){\rule[-0.200pt]{0.400pt}{8.191pt}}
\put(453.0,641.0){\rule[-0.200pt]{4.818pt}{0.400pt}}
\put(1591,511){\circle*{12}}
\put(1241,581){\circle*{12}}
\put(1025,697){\circle*{12}}
\put(777,653){\circle*{12}}
\put(608,664){\circle*{12}}
\put(517,625){\circle*{12}}
\put(463,658){\circle*{12}}
\put(453.0,675.0){\rule[-0.200pt]{4.818pt}{0.400pt}}
\put(375,656){\usebox{\plotpoint}}
\put(389,654.67){\rule{3.132pt}{0.400pt}}
\multiput(389.00,655.17)(6.500,-1.000){2}{\rule{1.566pt}{0.400pt}}
\put(402,653.67){\rule{3.373pt}{0.400pt}}
\multiput(402.00,654.17)(7.000,-1.000){2}{\rule{1.686pt}{0.400pt}}
\put(416,652.67){\rule{3.373pt}{0.400pt}}
\multiput(416.00,653.17)(7.000,-1.000){2}{\rule{1.686pt}{0.400pt}}
\put(375.0,656.0){\rule[-0.200pt]{3.373pt}{0.400pt}}
\put(443,651.67){\rule{3.373pt}{0.400pt}}
\multiput(443.00,652.17)(7.000,-1.000){2}{\rule{1.686pt}{0.400pt}}
\put(457,650.67){\rule{3.132pt}{0.400pt}}
\multiput(457.00,651.17)(6.500,-1.000){2}{\rule{1.566pt}{0.400pt}}
\put(470,649.67){\rule{3.373pt}{0.400pt}}
\multiput(470.00,650.17)(7.000,-1.000){2}{\rule{1.686pt}{0.400pt}}
\put(430.0,653.0){\rule[-0.200pt]{3.132pt}{0.400pt}}
\put(498,648.67){\rule{3.132pt}{0.400pt}}
\multiput(498.00,649.17)(6.500,-1.000){2}{\rule{1.566pt}{0.400pt}}
\put(511,647.67){\rule{3.373pt}{0.400pt}}
\multiput(511.00,648.17)(7.000,-1.000){2}{\rule{1.686pt}{0.400pt}}
\put(525,646.67){\rule{3.373pt}{0.400pt}}
\multiput(525.00,647.17)(7.000,-1.000){2}{\rule{1.686pt}{0.400pt}}
\put(539,645.67){\rule{3.132pt}{0.400pt}}
\multiput(539.00,646.17)(6.500,-1.000){2}{\rule{1.566pt}{0.400pt}}
\put(484.0,650.0){\rule[-0.200pt]{3.373pt}{0.400pt}}
\put(566,644.67){\rule{3.373pt}{0.400pt}}
\multiput(566.00,645.17)(7.000,-1.000){2}{\rule{1.686pt}{0.400pt}}
\put(580,643.67){\rule{3.132pt}{0.400pt}}
\multiput(580.00,644.17)(6.500,-1.000){2}{\rule{1.566pt}{0.400pt}}
\put(593,642.67){\rule{3.373pt}{0.400pt}}
\multiput(593.00,643.17)(7.000,-1.000){2}{\rule{1.686pt}{0.400pt}}
\put(552.0,646.0){\rule[-0.200pt]{3.373pt}{0.400pt}}
\put(620,641.67){\rule{3.373pt}{0.400pt}}
\multiput(620.00,642.17)(7.000,-1.000){2}{\rule{1.686pt}{0.400pt}}
\put(634,640.67){\rule{3.373pt}{0.400pt}}
\multiput(634.00,641.17)(7.000,-1.000){2}{\rule{1.686pt}{0.400pt}}
\put(648,639.67){\rule{3.132pt}{0.400pt}}
\multiput(648.00,640.17)(6.500,-1.000){2}{\rule{1.566pt}{0.400pt}}
\put(607.0,643.0){\rule[-0.200pt]{3.132pt}{0.400pt}}
\put(675,638.67){\rule{3.373pt}{0.400pt}}
\multiput(675.00,639.17)(7.000,-1.000){2}{\rule{1.686pt}{0.400pt}}
\put(689,637.67){\rule{3.132pt}{0.400pt}}
\multiput(689.00,638.17)(6.500,-1.000){2}{\rule{1.566pt}{0.400pt}}
\put(702,636.67){\rule{3.373pt}{0.400pt}}
\multiput(702.00,637.17)(7.000,-1.000){2}{\rule{1.686pt}{0.400pt}}
\put(661.0,640.0){\rule[-0.200pt]{3.373pt}{0.400pt}}
\put(730,635.67){\rule{3.132pt}{0.400pt}}
\multiput(730.00,636.17)(6.500,-1.000){2}{\rule{1.566pt}{0.400pt}}
\put(743,634.67){\rule{3.373pt}{0.400pt}}
\multiput(743.00,635.17)(7.000,-1.000){2}{\rule{1.686pt}{0.400pt}}
\put(757,633.67){\rule{3.132pt}{0.400pt}}
\multiput(757.00,634.17)(6.500,-1.000){2}{\rule{1.566pt}{0.400pt}}
\put(770,632.67){\rule{3.373pt}{0.400pt}}
\multiput(770.00,633.17)(7.000,-1.000){2}{\rule{1.686pt}{0.400pt}}
\put(716.0,637.0){\rule[-0.200pt]{3.373pt}{0.400pt}}
\put(798,631.67){\rule{3.132pt}{0.400pt}}
\multiput(798.00,632.17)(6.500,-1.000){2}{\rule{1.566pt}{0.400pt}}
\put(811,630.67){\rule{3.373pt}{0.400pt}}
\multiput(811.00,631.17)(7.000,-1.000){2}{\rule{1.686pt}{0.400pt}}
\put(825,629.67){\rule{3.373pt}{0.400pt}}
\multiput(825.00,630.17)(7.000,-1.000){2}{\rule{1.686pt}{0.400pt}}
\put(784.0,633.0){\rule[-0.200pt]{3.373pt}{0.400pt}}
\put(852,628.67){\rule{3.373pt}{0.400pt}}
\multiput(852.00,629.17)(7.000,-1.000){2}{\rule{1.686pt}{0.400pt}}
\put(866,627.67){\rule{3.373pt}{0.400pt}}
\multiput(866.00,628.17)(7.000,-1.000){2}{\rule{1.686pt}{0.400pt}}
\put(880,626.67){\rule{3.132pt}{0.400pt}}
\multiput(880.00,627.17)(6.500,-1.000){2}{\rule{1.566pt}{0.400pt}}
\put(839.0,630.0){\rule[-0.200pt]{3.132pt}{0.400pt}}
\put(907,625.67){\rule{3.132pt}{0.400pt}}
\multiput(907.00,626.17)(6.500,-1.000){2}{\rule{1.566pt}{0.400pt}}
\put(920,624.67){\rule{3.373pt}{0.400pt}}
\multiput(920.00,625.17)(7.000,-1.000){2}{\rule{1.686pt}{0.400pt}}
\put(934,623.67){\rule{3.373pt}{0.400pt}}
\multiput(934.00,624.17)(7.000,-1.000){2}{\rule{1.686pt}{0.400pt}}
\put(893.0,627.0){\rule[-0.200pt]{3.373pt}{0.400pt}}
\put(961,622.67){\rule{3.373pt}{0.400pt}}
\multiput(961.00,623.17)(7.000,-1.000){2}{\rule{1.686pt}{0.400pt}}
\put(975,621.67){\rule{3.373pt}{0.400pt}}
\multiput(975.00,622.17)(7.000,-1.000){2}{\rule{1.686pt}{0.400pt}}
\put(989,620.67){\rule{3.132pt}{0.400pt}}
\multiput(989.00,621.17)(6.500,-1.000){2}{\rule{1.566pt}{0.400pt}}
\put(1002,619.67){\rule{3.373pt}{0.400pt}}
\multiput(1002.00,620.17)(7.000,-1.000){2}{\rule{1.686pt}{0.400pt}}
\put(948.0,624.0){\rule[-0.200pt]{3.132pt}{0.400pt}}
\put(1030,618.67){\rule{3.132pt}{0.400pt}}
\multiput(1030.00,619.17)(6.500,-1.000){2}{\rule{1.566pt}{0.400pt}}
\put(1043,617.67){\rule{3.373pt}{0.400pt}}
\multiput(1043.00,618.17)(7.000,-1.000){2}{\rule{1.686pt}{0.400pt}}
\put(1057,616.67){\rule{3.132pt}{0.400pt}}
\multiput(1057.00,617.17)(6.500,-1.000){2}{\rule{1.566pt}{0.400pt}}
\put(1016.0,620.0){\rule[-0.200pt]{3.373pt}{0.400pt}}
\put(1084,615.67){\rule{3.373pt}{0.400pt}}
\multiput(1084.00,616.17)(7.000,-1.000){2}{\rule{1.686pt}{0.400pt}}
\put(1098,614.67){\rule{3.132pt}{0.400pt}}
\multiput(1098.00,615.17)(6.500,-1.000){2}{\rule{1.566pt}{0.400pt}}
\put(1111,613.67){\rule{3.373pt}{0.400pt}}
\multiput(1111.00,614.17)(7.000,-1.000){2}{\rule{1.686pt}{0.400pt}}
\put(1070.0,617.0){\rule[-0.200pt]{3.373pt}{0.400pt}}
\put(1139,612.67){\rule{3.132pt}{0.400pt}}
\multiput(1139.00,613.17)(6.500,-1.000){2}{\rule{1.566pt}{0.400pt}}
\put(1152,611.67){\rule{3.373pt}{0.400pt}}
\multiput(1152.00,612.17)(7.000,-1.000){2}{\rule{1.686pt}{0.400pt}}
\put(1166,610.67){\rule{3.373pt}{0.400pt}}
\multiput(1166.00,611.17)(7.000,-1.000){2}{\rule{1.686pt}{0.400pt}}
\put(1125.0,614.0){\rule[-0.200pt]{3.373pt}{0.400pt}}
\put(1193,609.67){\rule{3.373pt}{0.400pt}}
\multiput(1193.00,610.17)(7.000,-1.000){2}{\rule{1.686pt}{0.400pt}}
\put(1207,608.67){\rule{3.132pt}{0.400pt}}
\multiput(1207.00,609.17)(6.500,-1.000){2}{\rule{1.566pt}{0.400pt}}
\put(1220,607.67){\rule{3.373pt}{0.400pt}}
\multiput(1220.00,608.17)(7.000,-1.000){2}{\rule{1.686pt}{0.400pt}}
\put(1234,606.67){\rule{3.373pt}{0.400pt}}
\multiput(1234.00,607.17)(7.000,-1.000){2}{\rule{1.686pt}{0.400pt}}
\put(1180.0,611.0){\rule[-0.200pt]{3.132pt}{0.400pt}}
\put(1261,605.67){\rule{3.373pt}{0.400pt}}
\multiput(1261.00,606.17)(7.000,-1.000){2}{\rule{1.686pt}{0.400pt}}
\put(1275,604.67){\rule{3.373pt}{0.400pt}}
\multiput(1275.00,605.17)(7.000,-1.000){2}{\rule{1.686pt}{0.400pt}}
\put(1289,603.67){\rule{3.132pt}{0.400pt}}
\multiput(1289.00,604.17)(6.500,-1.000){2}{\rule{1.566pt}{0.400pt}}
\put(1248.0,607.0){\rule[-0.200pt]{3.132pt}{0.400pt}}
\put(1316,602.67){\rule{3.373pt}{0.400pt}}
\multiput(1316.00,603.17)(7.000,-1.000){2}{\rule{1.686pt}{0.400pt}}
\put(1330,601.67){\rule{3.132pt}{0.400pt}}
\multiput(1330.00,602.17)(6.500,-1.000){2}{\rule{1.566pt}{0.400pt}}
\put(1343,600.67){\rule{3.373pt}{0.400pt}}
\multiput(1343.00,601.17)(7.000,-1.000){2}{\rule{1.686pt}{0.400pt}}
\put(1302.0,604.0){\rule[-0.200pt]{3.373pt}{0.400pt}}
\put(1370,599.67){\rule{3.373pt}{0.400pt}}
\multiput(1370.00,600.17)(7.000,-1.000){2}{\rule{1.686pt}{0.400pt}}
\put(1384,598.67){\rule{3.373pt}{0.400pt}}
\multiput(1384.00,599.17)(7.000,-1.000){2}{\rule{1.686pt}{0.400pt}}
\put(1398,597.67){\rule{3.132pt}{0.400pt}}
\multiput(1398.00,598.17)(6.500,-1.000){2}{\rule{1.566pt}{0.400pt}}
\put(1357.0,601.0){\rule[-0.200pt]{3.132pt}{0.400pt}}
\put(1425,596.67){\rule{3.373pt}{0.400pt}}
\multiput(1425.00,597.17)(7.000,-1.000){2}{\rule{1.686pt}{0.400pt}}
\put(1439,595.67){\rule{3.132pt}{0.400pt}}
\multiput(1439.00,596.17)(6.500,-1.000){2}{\rule{1.566pt}{0.400pt}}
\put(1452,594.67){\rule{3.373pt}{0.400pt}}
\multiput(1452.00,595.17)(7.000,-1.000){2}{\rule{1.686pt}{0.400pt}}
\put(1466,593.67){\rule{3.373pt}{0.400pt}}
\multiput(1466.00,594.17)(7.000,-1.000){2}{\rule{1.686pt}{0.400pt}}
\put(1411.0,598.0){\rule[-0.200pt]{3.373pt}{0.400pt}}
\put(1493,592.67){\rule{3.373pt}{0.400pt}}
\multiput(1493.00,593.17)(7.000,-1.000){2}{\rule{1.686pt}{0.400pt}}
\put(1507,591.67){\rule{3.132pt}{0.400pt}}
\multiput(1507.00,592.17)(6.500,-1.000){2}{\rule{1.566pt}{0.400pt}}
\put(1520,590.67){\rule{3.373pt}{0.400pt}}
\multiput(1520.00,591.17)(7.000,-1.000){2}{\rule{1.686pt}{0.400pt}}
\put(1480.0,594.0){\rule[-0.200pt]{3.132pt}{0.400pt}}
\put(1548,589.67){\rule{3.132pt}{0.400pt}}
\multiput(1548.00,590.17)(6.500,-1.000){2}{\rule{1.566pt}{0.400pt}}
\put(1561,588.67){\rule{3.373pt}{0.400pt}}
\multiput(1561.00,589.17)(7.000,-1.000){2}{\rule{1.686pt}{0.400pt}}
\put(1575,587.67){\rule{3.373pt}{0.400pt}}
\multiput(1575.00,588.17)(7.000,-1.000){2}{\rule{1.686pt}{0.400pt}}
\put(1534.0,591.0){\rule[-0.200pt]{3.373pt}{0.400pt}}
\put(1602,586.67){\rule{3.373pt}{0.400pt}}
\multiput(1602.00,587.17)(7.000,-1.000){2}{\rule{1.686pt}{0.400pt}}
\put(1616,585.67){\rule{3.373pt}{0.400pt}}
\multiput(1616.00,586.17)(7.000,-1.000){2}{\rule{1.686pt}{0.400pt}}
\put(1630,584.67){\rule{3.132pt}{0.400pt}}
\multiput(1630.00,585.17)(6.500,-1.000){2}{\rule{1.566pt}{0.400pt}}
\put(1589.0,588.0){\rule[-0.200pt]{3.132pt}{0.400pt}}
\put(1657,583.67){\rule{3.132pt}{0.400pt}}
\multiput(1657.00,584.17)(6.500,-1.000){2}{\rule{1.566pt}{0.400pt}}
\put(1670,582.67){\rule{3.373pt}{0.400pt}}
\multiput(1670.00,583.17)(7.000,-1.000){2}{\rule{1.686pt}{0.400pt}}
\put(1684,581.67){\rule{3.373pt}{0.400pt}}
\multiput(1684.00,582.17)(7.000,-1.000){2}{\rule{1.686pt}{0.400pt}}
\put(1643.0,585.0){\rule[-0.200pt]{3.373pt}{0.400pt}}
\put(1711,580.67){\rule{3.373pt}{0.400pt}}
\multiput(1711.00,581.17)(7.000,-1.000){2}{\rule{1.686pt}{0.400pt}}
\put(1698.0,582.0){\rule[-0.200pt]{3.132pt}{0.400pt}}
\sbox{\plotpoint}{\rule[-0.500pt]{1.000pt}{1.000pt}}%
\put(375,624){\usebox{\plotpoint}}
\put(375.00,624.00){\usebox{\plotpoint}}
\put(395.36,627.98){\usebox{\plotpoint}}
\put(415.90,630.99){\usebox{\plotpoint}}
\put(436.26,634.96){\usebox{\plotpoint}}
\put(456.80,637.97){\usebox{\plotpoint}}
\put(477.32,641.05){\usebox{\plotpoint}}
\put(497.87,643.98){\usebox{\plotpoint}}
\put(518.51,646.07){\usebox{\plotpoint}}
\put(539.06,649.00){\usebox{\plotpoint}}
\put(559.76,650.55){\usebox{\plotpoint}}
\put(580.35,653.03){\usebox{\plotpoint}}
\put(601.05,654.57){\usebox{\plotpoint}}
\put(621.75,656.12){\usebox{\plotpoint}}
\put(642.45,657.60){\usebox{\plotpoint}}
\put(663.19,658.16){\usebox{\plotpoint}}
\put(683.89,659.63){\usebox{\plotpoint}}
\put(704.62,660.19){\usebox{\plotpoint}}
\put(725.35,661.00){\usebox{\plotpoint}}
\put(746.11,661.00){\usebox{\plotpoint}}
\put(766.86,661.00){\usebox{\plotpoint}}
\put(787.62,661.00){\usebox{\plotpoint}}
\put(808.37,661.00){\usebox{\plotpoint}}
\put(829.09,660.00){\usebox{\plotpoint}}
\put(849.82,659.17){\usebox{\plotpoint}}
\put(870.55,658.67){\usebox{\plotpoint}}
\put(891.25,657.13){\usebox{\plotpoint}}
\put(911.99,656.62){\usebox{\plotpoint}}
\put(932.69,655.09){\usebox{\plotpoint}}
\put(953.28,652.59){\usebox{\plotpoint}}
\put(973.98,651.07){\usebox{\plotpoint}}
\put(994.58,648.57){\usebox{\plotpoint}}
\put(1015.18,646.12){\usebox{\plotpoint}}
\put(1035.82,644.10){\usebox{\plotpoint}}
\put(1056.35,641.09){\usebox{\plotpoint}}
\put(1076.88,638.02){\usebox{\plotpoint}}
\put(1097.43,635.08){\usebox{\plotpoint}}
\put(1117.87,631.53){\usebox{\plotpoint}}
\put(1138.33,628.10){\usebox{\plotpoint}}
\put(1158.77,624.55){\usebox{\plotpoint}}
\put(1179.06,620.20){\usebox{\plotpoint}}
\put(1199.50,616.61){\usebox{\plotpoint}}
\put(1219.75,612.06){\usebox{\plotpoint}}
\put(1240.04,607.71){\usebox{\plotpoint}}
\put(1260.29,603.16){\usebox{\plotpoint}}
\put(1280.35,597.85){\usebox{\plotpoint}}
\put(1300.38,592.50){\usebox{\plotpoint}}
\put(1320.56,587.70){\usebox{\plotpoint}}
\put(1340.66,582.54){\usebox{\plotpoint}}
\put(1360.62,576.88){\usebox{\plotpoint}}
\put(1380.52,570.99){\usebox{\plotpoint}}
\put(1400.47,565.24){\usebox{\plotpoint}}
\put(1420.17,558.73){\usebox{\plotpoint}}
\put(1439.99,552.62){\usebox{\plotpoint}}
\put(1459.59,545.83){\usebox{\plotpoint}}
\put(1479.27,539.26){\usebox{\plotpoint}}
\put(1499.00,532.86){\usebox{\plotpoint}}
\put(1518.45,525.60){\usebox{\plotpoint}}
\put(1537.98,518.58){\usebox{\plotpoint}}
\put(1557.44,511.37){\usebox{\plotpoint}}
\put(1576.61,503.42){\usebox{\plotpoint}}
\put(1596.09,496.27){\usebox{\plotpoint}}
\put(1615.26,488.32){\usebox{\plotpoint}}
\put(1634.62,480.87){\usebox{\plotpoint}}
\put(1653.59,472.46){\usebox{\plotpoint}}
\put(1672.51,463.93){\usebox{\plotpoint}}
\put(1691.58,455.75){\usebox{\plotpoint}}
\put(1710.51,447.23){\usebox{\plotpoint}}
\put(1725,441){\usebox{\plotpoint}}
\end{picture}

\end	{center}
\caption{$k=2$ string tension in SU(5), extracted using single
exponential, $S$, fits, as a 
function of the lattice spacing. Continuum extrapolations linear and
quadratic in $a^2$ are shown.}
\label{fig_n5Scont}
\end 	{figure}
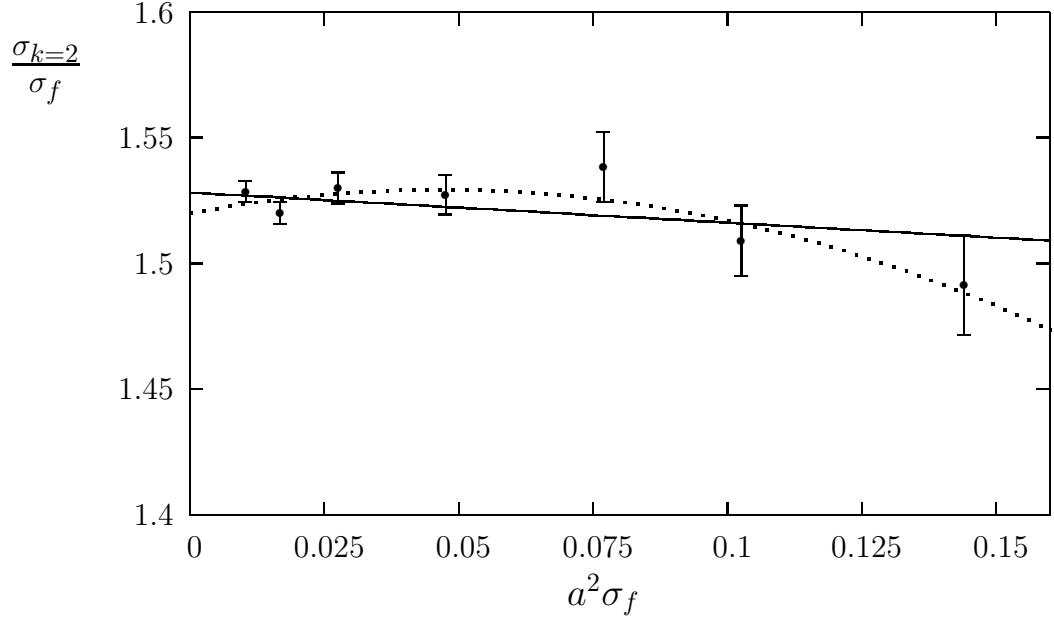

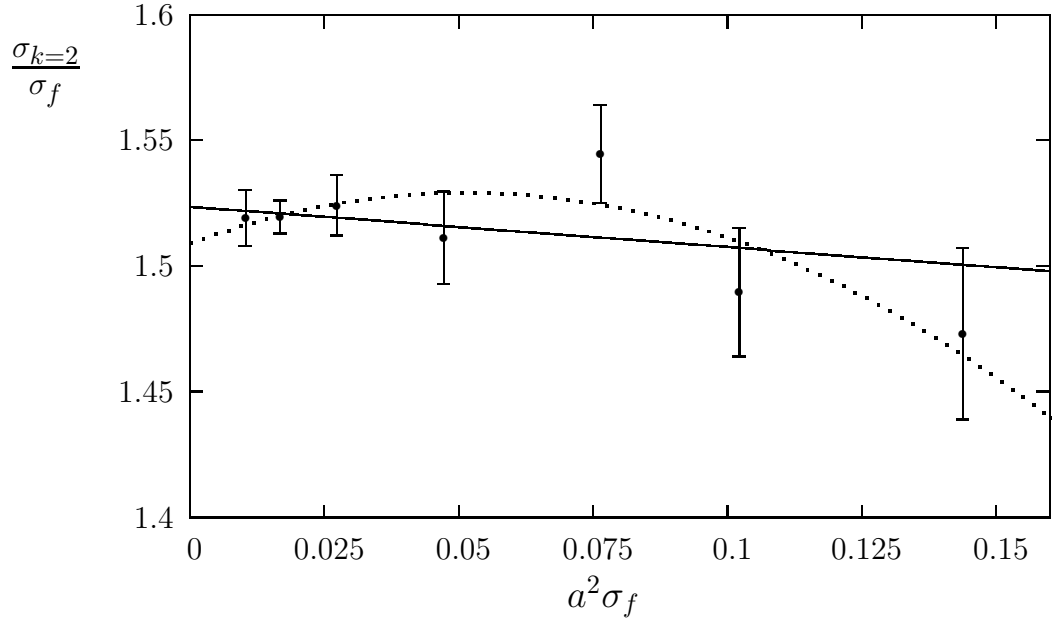
\begin	{figure}[p]
\begin	{center}
\leavevmode
\setlength{\unitlength}{0.240900pt}
\ifx\plotpoint\undefined\newsavebox{\plotpoint}\fi
\sbox{\plotpoint}{\rule[-0.200pt]{0.400pt}{0.400pt}}%
\begin{picture}(1800,990)(0,0)
\font\gnuplot=cmr10 at 12pt
\gnuplot
\sbox{\plotpoint}{\rule[-0.200pt]{0.400pt}{0.400pt}}%
\put(375.0,150.0){\rule[-0.200pt]{4.818pt}{0.400pt}}
\put(350,150){\makebox(0,0)[r]{\ \ {$1.4$}}}
\put(1705.0,150.0){\rule[-0.200pt]{4.818pt}{0.400pt}}
\put(375.0,348.0){\rule[-0.200pt]{4.818pt}{0.400pt}}
\put(350,348){\makebox(0,0)[r]{\ \ {$1.45$}}}
\put(1705.0,348.0){\rule[-0.200pt]{4.818pt}{0.400pt}}
\put(375.0,545.0){\rule[-0.200pt]{4.818pt}{0.400pt}}
\put(350,545){\makebox(0,0)[r]{\ \ {$1.5$}}}
\put(1705.0,545.0){\rule[-0.200pt]{4.818pt}{0.400pt}}
\put(375.0,743.0){\rule[-0.200pt]{4.818pt}{0.400pt}}
\put(350,743){\makebox(0,0)[r]{\ \ {$1.55$}}}
\put(1705.0,743.0){\rule[-0.200pt]{4.818pt}{0.400pt}}
\put(375.0,940.0){\rule[-0.200pt]{4.818pt}{0.400pt}}
\put(350,940){\makebox(0,0)[r]{\ \ {$1.6$}}}
\put(1705.0,940.0){\rule[-0.200pt]{4.818pt}{0.400pt}}
\put(375.0,150.0){\rule[-0.200pt]{0.400pt}{4.818pt}}
\put(375,100){\makebox(0,0){\ {$0$}}}
\put(375.0,920.0){\rule[-0.200pt]{0.400pt}{4.818pt}}
\put(586.0,150.0){\rule[-0.200pt]{0.400pt}{4.818pt}}
\put(586,100){\makebox(0,0){\ {$0.025$}}}
\put(586.0,920.0){\rule[-0.200pt]{0.400pt}{4.818pt}}
\put(797.0,150.0){\rule[-0.200pt]{0.400pt}{4.818pt}}
\put(797,100){\makebox(0,0){\ {$0.05$}}}
\put(797.0,920.0){\rule[-0.200pt]{0.400pt}{4.818pt}}
\put(1008.0,150.0){\rule[-0.200pt]{0.400pt}{4.818pt}}
\put(1008,100){\makebox(0,0){\ {$0.075$}}}
\put(1008.0,920.0){\rule[-0.200pt]{0.400pt}{4.818pt}}
\put(1219.0,150.0){\rule[-0.200pt]{0.400pt}{4.818pt}}
\put(1219,100){\makebox(0,0){\ {$0.1$}}}
\put(1219.0,920.0){\rule[-0.200pt]{0.400pt}{4.818pt}}
\put(1430.0,150.0){\rule[-0.200pt]{0.400pt}{4.818pt}}
\put(1430,100){\makebox(0,0){\ {$0.125$}}}
\put(1430.0,920.0){\rule[-0.200pt]{0.400pt}{4.818pt}}
\put(1641.0,150.0){\rule[-0.200pt]{0.400pt}{4.818pt}}
\put(1641,100){\makebox(0,0){\ {$0.15$}}}
\put(1641.0,920.0){\rule[-0.200pt]{0.400pt}{4.818pt}}
\put(375.0,150.0){\rule[-0.200pt]{325.215pt}{0.400pt}}
\put(1725.0,150.0){\rule[-0.200pt]{0.400pt}{190.311pt}}
\put(375.0,940.0){\rule[-0.200pt]{325.215pt}{0.400pt}}
\put(150,845){\makebox(0,0){\Large{${{\sigma_{k=2}}\over{\sigma_f}}$}}}
\put(1025,25){\makebox(0,0){\large{$a^2\sigma_f$}}}
\put(375.0,150.0){\rule[-0.200pt]{0.400pt}{190.311pt}}
\put(1589.0,304.0){\rule[-0.200pt]{0.400pt}{64.802pt}}
\put(1579.0,304.0){\rule[-0.200pt]{4.818pt}{0.400pt}}
\put(1579.0,573.0){\rule[-0.200pt]{4.818pt}{0.400pt}}
\put(1238.0,403.0){\rule[-0.200pt]{0.400pt}{48.662pt}}
\put(1228.0,403.0){\rule[-0.200pt]{4.818pt}{0.400pt}}
\put(1228.0,605.0){\rule[-0.200pt]{4.818pt}{0.400pt}}
\put(1020.0,644.0){\rule[-0.200pt]{0.400pt}{37.099pt}}
\put(1010.0,644.0){\rule[-0.200pt]{4.818pt}{0.400pt}}
\put(1010.0,798.0){\rule[-0.200pt]{4.818pt}{0.400pt}}
\put(774.0,517.0){\rule[-0.200pt]{0.400pt}{34.930pt}}
\put(764.0,517.0){\rule[-0.200pt]{4.818pt}{0.400pt}}
\put(764.0,662.0){\rule[-0.200pt]{4.818pt}{0.400pt}}
\put(606.0,593.0){\rule[-0.200pt]{0.400pt}{22.885pt}}
\put(596.0,593.0){\rule[-0.200pt]{4.818pt}{0.400pt}}
\put(596.0,688.0){\rule[-0.200pt]{4.818pt}{0.400pt}}
\put(517.0,596.0){\rule[-0.200pt]{0.400pt}{12.527pt}}
\put(507.0,596.0){\rule[-0.200pt]{4.818pt}{0.400pt}}
\put(507.0,648.0){\rule[-0.200pt]{4.818pt}{0.400pt}}
\put(463.0,576.0){\rule[-0.200pt]{0.400pt}{21.199pt}}
\put(453.0,576.0){\rule[-0.200pt]{4.818pt}{0.400pt}}
\put(1589,438){\circle*{12}}
\put(1238,504){\circle*{12}}
\put(1020,721){\circle*{12}}
\put(774,589){\circle*{12}}
\put(606,640){\circle*{12}}
\put(517,622){\circle*{12}}
\put(463,620){\circle*{12}}
\put(453.0,664.0){\rule[-0.200pt]{4.818pt}{0.400pt}}
\put(375,638){\usebox{\plotpoint}}
\put(375,636.67){\rule{3.373pt}{0.400pt}}
\multiput(375.00,637.17)(7.000,-1.000){2}{\rule{1.686pt}{0.400pt}}
\put(389,635.67){\rule{3.132pt}{0.400pt}}
\multiput(389.00,636.17)(6.500,-1.000){2}{\rule{1.566pt}{0.400pt}}
\put(402,634.67){\rule{3.373pt}{0.400pt}}
\multiput(402.00,635.17)(7.000,-1.000){2}{\rule{1.686pt}{0.400pt}}
\put(416,633.67){\rule{3.373pt}{0.400pt}}
\multiput(416.00,634.17)(7.000,-1.000){2}{\rule{1.686pt}{0.400pt}}
\put(430,632.67){\rule{3.132pt}{0.400pt}}
\multiput(430.00,633.17)(6.500,-1.000){2}{\rule{1.566pt}{0.400pt}}
\put(443,631.67){\rule{3.373pt}{0.400pt}}
\multiput(443.00,632.17)(7.000,-1.000){2}{\rule{1.686pt}{0.400pt}}
\put(457,630.67){\rule{3.132pt}{0.400pt}}
\multiput(457.00,631.17)(6.500,-1.000){2}{\rule{1.566pt}{0.400pt}}
\put(470,629.67){\rule{3.373pt}{0.400pt}}
\multiput(470.00,630.17)(7.000,-1.000){2}{\rule{1.686pt}{0.400pt}}
\put(484,628.67){\rule{3.373pt}{0.400pt}}
\multiput(484.00,629.17)(7.000,-1.000){2}{\rule{1.686pt}{0.400pt}}
\put(498,627.67){\rule{3.132pt}{0.400pt}}
\multiput(498.00,628.17)(6.500,-1.000){2}{\rule{1.566pt}{0.400pt}}
\put(511,626.67){\rule{3.373pt}{0.400pt}}
\multiput(511.00,627.17)(7.000,-1.000){2}{\rule{1.686pt}{0.400pt}}
\put(525,625.67){\rule{3.373pt}{0.400pt}}
\multiput(525.00,626.17)(7.000,-1.000){2}{\rule{1.686pt}{0.400pt}}
\put(539,624.67){\rule{3.132pt}{0.400pt}}
\multiput(539.00,625.17)(6.500,-1.000){2}{\rule{1.566pt}{0.400pt}}
\put(552,623.67){\rule{3.373pt}{0.400pt}}
\multiput(552.00,624.17)(7.000,-1.000){2}{\rule{1.686pt}{0.400pt}}
\put(566,622.67){\rule{3.373pt}{0.400pt}}
\multiput(566.00,623.17)(7.000,-1.000){2}{\rule{1.686pt}{0.400pt}}
\put(580,621.67){\rule{3.132pt}{0.400pt}}
\multiput(580.00,622.17)(6.500,-1.000){2}{\rule{1.566pt}{0.400pt}}
\put(593,620.67){\rule{3.373pt}{0.400pt}}
\multiput(593.00,621.17)(7.000,-1.000){2}{\rule{1.686pt}{0.400pt}}
\put(607,619.67){\rule{3.132pt}{0.400pt}}
\multiput(607.00,620.17)(6.500,-1.000){2}{\rule{1.566pt}{0.400pt}}
\put(620,618.67){\rule{3.373pt}{0.400pt}}
\multiput(620.00,619.17)(7.000,-1.000){2}{\rule{1.686pt}{0.400pt}}
\put(634,617.67){\rule{3.373pt}{0.400pt}}
\multiput(634.00,618.17)(7.000,-1.000){2}{\rule{1.686pt}{0.400pt}}
\put(648,616.67){\rule{3.132pt}{0.400pt}}
\multiput(648.00,617.17)(6.500,-1.000){2}{\rule{1.566pt}{0.400pt}}
\put(661,615.67){\rule{3.373pt}{0.400pt}}
\multiput(661.00,616.17)(7.000,-1.000){2}{\rule{1.686pt}{0.400pt}}
\put(675,614.67){\rule{3.373pt}{0.400pt}}
\multiput(675.00,615.17)(7.000,-1.000){2}{\rule{1.686pt}{0.400pt}}
\put(689,613.67){\rule{3.132pt}{0.400pt}}
\multiput(689.00,614.17)(6.500,-1.000){2}{\rule{1.566pt}{0.400pt}}
\put(702,612.67){\rule{3.373pt}{0.400pt}}
\multiput(702.00,613.17)(7.000,-1.000){2}{\rule{1.686pt}{0.400pt}}
\put(716,611.67){\rule{3.373pt}{0.400pt}}
\multiput(716.00,612.17)(7.000,-1.000){2}{\rule{1.686pt}{0.400pt}}
\put(730,610.67){\rule{3.132pt}{0.400pt}}
\multiput(730.00,611.17)(6.500,-1.000){2}{\rule{1.566pt}{0.400pt}}
\put(743,609.17){\rule{2.900pt}{0.400pt}}
\multiput(743.00,610.17)(7.981,-2.000){2}{\rule{1.450pt}{0.400pt}}
\put(757,607.67){\rule{3.132pt}{0.400pt}}
\multiput(757.00,608.17)(6.500,-1.000){2}{\rule{1.566pt}{0.400pt}}
\put(770,606.67){\rule{3.373pt}{0.400pt}}
\multiput(770.00,607.17)(7.000,-1.000){2}{\rule{1.686pt}{0.400pt}}
\put(784,605.67){\rule{3.373pt}{0.400pt}}
\multiput(784.00,606.17)(7.000,-1.000){2}{\rule{1.686pt}{0.400pt}}
\put(798,604.67){\rule{3.132pt}{0.400pt}}
\multiput(798.00,605.17)(6.500,-1.000){2}{\rule{1.566pt}{0.400pt}}
\put(811,603.67){\rule{3.373pt}{0.400pt}}
\multiput(811.00,604.17)(7.000,-1.000){2}{\rule{1.686pt}{0.400pt}}
\put(825,602.67){\rule{3.373pt}{0.400pt}}
\multiput(825.00,603.17)(7.000,-1.000){2}{\rule{1.686pt}{0.400pt}}
\put(839,601.67){\rule{3.132pt}{0.400pt}}
\multiput(839.00,602.17)(6.500,-1.000){2}{\rule{1.566pt}{0.400pt}}
\put(852,600.67){\rule{3.373pt}{0.400pt}}
\multiput(852.00,601.17)(7.000,-1.000){2}{\rule{1.686pt}{0.400pt}}
\put(866,599.67){\rule{3.373pt}{0.400pt}}
\multiput(866.00,600.17)(7.000,-1.000){2}{\rule{1.686pt}{0.400pt}}
\put(880,598.67){\rule{3.132pt}{0.400pt}}
\multiput(880.00,599.17)(6.500,-1.000){2}{\rule{1.566pt}{0.400pt}}
\put(893,597.67){\rule{3.373pt}{0.400pt}}
\multiput(893.00,598.17)(7.000,-1.000){2}{\rule{1.686pt}{0.400pt}}
\put(907,596.67){\rule{3.132pt}{0.400pt}}
\multiput(907.00,597.17)(6.500,-1.000){2}{\rule{1.566pt}{0.400pt}}
\put(920,595.67){\rule{3.373pt}{0.400pt}}
\multiput(920.00,596.17)(7.000,-1.000){2}{\rule{1.686pt}{0.400pt}}
\put(934,594.67){\rule{3.373pt}{0.400pt}}
\multiput(934.00,595.17)(7.000,-1.000){2}{\rule{1.686pt}{0.400pt}}
\put(948,593.67){\rule{3.132pt}{0.400pt}}
\multiput(948.00,594.17)(6.500,-1.000){2}{\rule{1.566pt}{0.400pt}}
\put(961,592.67){\rule{3.373pt}{0.400pt}}
\multiput(961.00,593.17)(7.000,-1.000){2}{\rule{1.686pt}{0.400pt}}
\put(975,591.67){\rule{3.373pt}{0.400pt}}
\multiput(975.00,592.17)(7.000,-1.000){2}{\rule{1.686pt}{0.400pt}}
\put(989,590.67){\rule{3.132pt}{0.400pt}}
\multiput(989.00,591.17)(6.500,-1.000){2}{\rule{1.566pt}{0.400pt}}
\put(1002,589.67){\rule{3.373pt}{0.400pt}}
\multiput(1002.00,590.17)(7.000,-1.000){2}{\rule{1.686pt}{0.400pt}}
\put(1016,588.67){\rule{3.373pt}{0.400pt}}
\multiput(1016.00,589.17)(7.000,-1.000){2}{\rule{1.686pt}{0.400pt}}
\put(1030,587.67){\rule{3.132pt}{0.400pt}}
\multiput(1030.00,588.17)(6.500,-1.000){2}{\rule{1.566pt}{0.400pt}}
\put(1043,586.67){\rule{3.373pt}{0.400pt}}
\multiput(1043.00,587.17)(7.000,-1.000){2}{\rule{1.686pt}{0.400pt}}
\put(1057,585.67){\rule{3.132pt}{0.400pt}}
\multiput(1057.00,586.17)(6.500,-1.000){2}{\rule{1.566pt}{0.400pt}}
\put(1070,584.67){\rule{3.373pt}{0.400pt}}
\multiput(1070.00,585.17)(7.000,-1.000){2}{\rule{1.686pt}{0.400pt}}
\put(1084,583.67){\rule{3.373pt}{0.400pt}}
\multiput(1084.00,584.17)(7.000,-1.000){2}{\rule{1.686pt}{0.400pt}}
\put(1098,582.67){\rule{3.132pt}{0.400pt}}
\multiput(1098.00,583.17)(6.500,-1.000){2}{\rule{1.566pt}{0.400pt}}
\put(1111,581.67){\rule{3.373pt}{0.400pt}}
\multiput(1111.00,582.17)(7.000,-1.000){2}{\rule{1.686pt}{0.400pt}}
\put(1125,580.67){\rule{3.373pt}{0.400pt}}
\multiput(1125.00,581.17)(7.000,-1.000){2}{\rule{1.686pt}{0.400pt}}
\put(1139,579.67){\rule{3.132pt}{0.400pt}}
\multiput(1139.00,580.17)(6.500,-1.000){2}{\rule{1.566pt}{0.400pt}}
\put(1152,578.67){\rule{3.373pt}{0.400pt}}
\multiput(1152.00,579.17)(7.000,-1.000){2}{\rule{1.686pt}{0.400pt}}
\put(1166,577.67){\rule{3.373pt}{0.400pt}}
\multiput(1166.00,578.17)(7.000,-1.000){2}{\rule{1.686pt}{0.400pt}}
\put(1180,576.67){\rule{3.132pt}{0.400pt}}
\multiput(1180.00,577.17)(6.500,-1.000){2}{\rule{1.566pt}{0.400pt}}
\put(1193,575.67){\rule{3.373pt}{0.400pt}}
\multiput(1193.00,576.17)(7.000,-1.000){2}{\rule{1.686pt}{0.400pt}}
\put(1207,574.67){\rule{3.132pt}{0.400pt}}
\multiput(1207.00,575.17)(6.500,-1.000){2}{\rule{1.566pt}{0.400pt}}
\put(1220,573.67){\rule{3.373pt}{0.400pt}}
\multiput(1220.00,574.17)(7.000,-1.000){2}{\rule{1.686pt}{0.400pt}}
\put(1234,572.67){\rule{3.373pt}{0.400pt}}
\multiput(1234.00,573.17)(7.000,-1.000){2}{\rule{1.686pt}{0.400pt}}
\put(1248,571.67){\rule{3.132pt}{0.400pt}}
\multiput(1248.00,572.17)(6.500,-1.000){2}{\rule{1.566pt}{0.400pt}}
\put(1261,570.67){\rule{3.373pt}{0.400pt}}
\multiput(1261.00,571.17)(7.000,-1.000){2}{\rule{1.686pt}{0.400pt}}
\put(1275,569.17){\rule{2.900pt}{0.400pt}}
\multiput(1275.00,570.17)(7.981,-2.000){2}{\rule{1.450pt}{0.400pt}}
\put(1289,567.67){\rule{3.132pt}{0.400pt}}
\multiput(1289.00,568.17)(6.500,-1.000){2}{\rule{1.566pt}{0.400pt}}
\put(1302,566.67){\rule{3.373pt}{0.400pt}}
\multiput(1302.00,567.17)(7.000,-1.000){2}{\rule{1.686pt}{0.400pt}}
\put(1316,565.67){\rule{3.373pt}{0.400pt}}
\multiput(1316.00,566.17)(7.000,-1.000){2}{\rule{1.686pt}{0.400pt}}
\put(1330,564.67){\rule{3.132pt}{0.400pt}}
\multiput(1330.00,565.17)(6.500,-1.000){2}{\rule{1.566pt}{0.400pt}}
\put(1343,563.67){\rule{3.373pt}{0.400pt}}
\multiput(1343.00,564.17)(7.000,-1.000){2}{\rule{1.686pt}{0.400pt}}
\put(1357,562.67){\rule{3.132pt}{0.400pt}}
\multiput(1357.00,563.17)(6.500,-1.000){2}{\rule{1.566pt}{0.400pt}}
\put(1370,561.67){\rule{3.373pt}{0.400pt}}
\multiput(1370.00,562.17)(7.000,-1.000){2}{\rule{1.686pt}{0.400pt}}
\put(1384,560.67){\rule{3.373pt}{0.400pt}}
\multiput(1384.00,561.17)(7.000,-1.000){2}{\rule{1.686pt}{0.400pt}}
\put(1398,559.67){\rule{3.132pt}{0.400pt}}
\multiput(1398.00,560.17)(6.500,-1.000){2}{\rule{1.566pt}{0.400pt}}
\put(1411,558.67){\rule{3.373pt}{0.400pt}}
\multiput(1411.00,559.17)(7.000,-1.000){2}{\rule{1.686pt}{0.400pt}}
\put(1425,557.67){\rule{3.373pt}{0.400pt}}
\multiput(1425.00,558.17)(7.000,-1.000){2}{\rule{1.686pt}{0.400pt}}
\put(1439,556.67){\rule{3.132pt}{0.400pt}}
\multiput(1439.00,557.17)(6.500,-1.000){2}{\rule{1.566pt}{0.400pt}}
\put(1452,555.67){\rule{3.373pt}{0.400pt}}
\multiput(1452.00,556.17)(7.000,-1.000){2}{\rule{1.686pt}{0.400pt}}
\put(1466,554.67){\rule{3.373pt}{0.400pt}}
\multiput(1466.00,555.17)(7.000,-1.000){2}{\rule{1.686pt}{0.400pt}}
\put(1480,553.67){\rule{3.132pt}{0.400pt}}
\multiput(1480.00,554.17)(6.500,-1.000){2}{\rule{1.566pt}{0.400pt}}
\put(1493,552.67){\rule{3.373pt}{0.400pt}}
\multiput(1493.00,553.17)(7.000,-1.000){2}{\rule{1.686pt}{0.400pt}}
\put(1507,551.67){\rule{3.132pt}{0.400pt}}
\multiput(1507.00,552.17)(6.500,-1.000){2}{\rule{1.566pt}{0.400pt}}
\put(1520,550.67){\rule{3.373pt}{0.400pt}}
\multiput(1520.00,551.17)(7.000,-1.000){2}{\rule{1.686pt}{0.400pt}}
\put(1534,549.67){\rule{3.373pt}{0.400pt}}
\multiput(1534.00,550.17)(7.000,-1.000){2}{\rule{1.686pt}{0.400pt}}
\put(1548,548.67){\rule{3.132pt}{0.400pt}}
\multiput(1548.00,549.17)(6.500,-1.000){2}{\rule{1.566pt}{0.400pt}}
\put(1561,547.67){\rule{3.373pt}{0.400pt}}
\multiput(1561.00,548.17)(7.000,-1.000){2}{\rule{1.686pt}{0.400pt}}
\put(1575,546.67){\rule{3.373pt}{0.400pt}}
\multiput(1575.00,547.17)(7.000,-1.000){2}{\rule{1.686pt}{0.400pt}}
\put(1589,545.67){\rule{3.132pt}{0.400pt}}
\multiput(1589.00,546.17)(6.500,-1.000){2}{\rule{1.566pt}{0.400pt}}
\put(1602,544.67){\rule{3.373pt}{0.400pt}}
\multiput(1602.00,545.17)(7.000,-1.000){2}{\rule{1.686pt}{0.400pt}}
\put(1616,543.67){\rule{3.373pt}{0.400pt}}
\multiput(1616.00,544.17)(7.000,-1.000){2}{\rule{1.686pt}{0.400pt}}
\put(1630,542.67){\rule{3.132pt}{0.400pt}}
\multiput(1630.00,543.17)(6.500,-1.000){2}{\rule{1.566pt}{0.400pt}}
\put(1643,541.67){\rule{3.373pt}{0.400pt}}
\multiput(1643.00,542.17)(7.000,-1.000){2}{\rule{1.686pt}{0.400pt}}
\put(1657,540.67){\rule{3.132pt}{0.400pt}}
\multiput(1657.00,541.17)(6.500,-1.000){2}{\rule{1.566pt}{0.400pt}}
\put(1670,539.67){\rule{3.373pt}{0.400pt}}
\multiput(1670.00,540.17)(7.000,-1.000){2}{\rule{1.686pt}{0.400pt}}
\put(1684,538.67){\rule{3.373pt}{0.400pt}}
\multiput(1684.00,539.17)(7.000,-1.000){2}{\rule{1.686pt}{0.400pt}}
\put(1698,537.67){\rule{3.132pt}{0.400pt}}
\multiput(1698.00,538.17)(6.500,-1.000){2}{\rule{1.566pt}{0.400pt}}
\put(1711,536.67){\rule{3.373pt}{0.400pt}}
\multiput(1711.00,537.17)(7.000,-1.000){2}{\rule{1.686pt}{0.400pt}}
\sbox{\plotpoint}{\rule[-0.500pt]{1.000pt}{1.000pt}}%
\put(375,581){\usebox{\plotpoint}}
\put(375.00,581.00){\usebox{\plotpoint}}
\put(394.78,587.22){\usebox{\plotpoint}}
\put(414.26,594.38){\usebox{\plotpoint}}
\put(434.16,600.28){\usebox{\plotpoint}}
\put(453.84,606.87){\usebox{\plotpoint}}
\put(473.71,612.79){\usebox{\plotpoint}}
\put(493.84,617.81){\usebox{\plotpoint}}
\put(513.76,623.59){\usebox{\plotpoint}}
\put(534.06,627.94){\usebox{\plotpoint}}
\put(554.31,632.49){\usebox{\plotpoint}}
\put(574.60,636.84){\usebox{\plotpoint}}
\put(595.04,640.44){\usebox{\plotpoint}}
\put(615.42,644.30){\usebox{\plotpoint}}
\put(635.96,647.28){\usebox{\plotpoint}}
\put(656.49,650.31){\usebox{\plotpoint}}
\put(677.14,652.31){\usebox{\plotpoint}}
\put(697.75,654.67){\usebox{\plotpoint}}
\put(718.45,656.17){\usebox{\plotpoint}}
\put(739.15,657.70){\usebox{\plotpoint}}
\put(759.86,659.00){\usebox{\plotpoint}}
\put(780.61,659.00){\usebox{\plotpoint}}
\put(801.33,660.00){\usebox{\plotpoint}}
\put(822.09,660.00){\usebox{\plotpoint}}
\put(842.81,659.00){\usebox{\plotpoint}}
\put(863.53,658.18){\usebox{\plotpoint}}
\put(884.27,657.67){\usebox{\plotpoint}}
\put(904.97,656.15){\usebox{\plotpoint}}
\put(925.55,653.60){\usebox{\plotpoint}}
\put(946.25,652.12){\usebox{\plotpoint}}
\put(966.79,649.17){\usebox{\plotpoint}}
\put(987.34,646.24){\usebox{\plotpoint}}
\put(1007.87,643.16){\usebox{\plotpoint}}
\put(1028.26,639.37){\usebox{\plotpoint}}
\put(1048.70,635.78){\usebox{\plotpoint}}
\put(1068.95,631.24){\usebox{\plotpoint}}
\put(1089.24,626.88){\usebox{\plotpoint}}
\put(1109.27,621.53){\usebox{\plotpoint}}
\put(1129.45,616.73){\usebox{\plotpoint}}
\put(1149.35,610.82){\usebox{\plotpoint}}
\put(1169.29,605.06){\usebox{\plotpoint}}
\put(1189.19,599.17){\usebox{\plotpoint}}
\put(1208.78,592.32){\usebox{\plotpoint}}
\put(1228.40,585.60){\usebox{\plotpoint}}
\put(1248.05,578.97){\usebox{\plotpoint}}
\put(1267.12,570.81){\usebox{\plotpoint}}
\put(1286.67,563.83){\usebox{\plotpoint}}
\put(1305.64,555.44){\usebox{\plotpoint}}
\put(1324.93,547.81){\usebox{\plotpoint}}
\put(1343.56,538.76){\usebox{\plotpoint}}
\put(1362.57,530.43){\usebox{\plotpoint}}
\put(1381.24,521.38){\usebox{\plotpoint}}
\put(1400.15,512.84){\usebox{\plotpoint}}
\put(1418.54,503.23){\usebox{\plotpoint}}
\put(1437.11,493.95){\usebox{\plotpoint}}
\put(1455.02,483.49){\usebox{\plotpoint}}
\put(1473.36,473.79){\usebox{\plotpoint}}
\put(1491.54,463.79){\usebox{\plotpoint}}
\put(1509.44,453.31){\usebox{\plotpoint}}
\put(1526.87,442.07){\usebox{\plotpoint}}
\put(1544.90,431.77){\usebox{\plotpoint}}
\put(1562.19,420.32){\usebox{\plotpoint}}
\put(1580.05,409.76){\usebox{\plotpoint}}
\put(1597.31,398.24){\usebox{\plotpoint}}
\put(1614.25,386.25){\usebox{\plotpoint}}
\put(1631.61,374.88){\usebox{\plotpoint}}
\put(1648.62,362.99){\usebox{\plotpoint}}
\put(1665.29,350.62){\usebox{\plotpoint}}
\put(1682.05,338.39){\usebox{\plotpoint}}
\put(1698.92,326.29){\usebox{\plotpoint}}
\put(1715.33,313.59){\usebox{\plotpoint}}
\put(1725,306){\usebox{\plotpoint}}
\end{picture}

\end	{center}
\caption{As in Fig.\ref{fig_n5Scont}, but using double exponential,
$D$, fits.}
\label{fig_n5Dcont}
\end 	{figure}

We show the resulting values of $\sigma_k/\sigma_f$ in SU(5)
in Figs.\ref{fig_n5Scont},\ref{fig_n5Dcont}, where the
calculations have been performed using single and
double exponential fits respectively. We also show 
linear $O(a^2)$ and quadratic $O(a^4)$ extrapolations to the 
continuum limit in each case. The resulting continuum values, 
together with the values of $\chi^2$ per degree of freedom for 
the best fits, are listed in Table~\ref{table_n5cont}.
Comparing Fig.\ref{fig_n5Scont} with Fig.\ref{fig_n5Dcont}
we see that the $S$ and $D$ values are quite consistent  
although the errors of the latter are significantly larger.
The corresponding continuum limits are, as a result, also
very similar. If we now compare the two different continuum
extrapolations, we see that including an extra $O(a^4)$ 
correction reduces the continuum value by a little more than 
one standard deviation. In principle the fit with the extra 
correction term should be more reliable. However the coefficient
of such a term is in practice largely determined by the calculations
at the largest values of $a$, where there is always the danger
that one is being  influenced by the strong-to-weak
coupling crossover where the power series expansion in $a^2$
breaks down. In any case, we see
from this study that neither the use of double exponential fits, 
nor the inclusion of higher order terms in the continuum
extrapolation, makes much difference to the final result.
Our qualitative conclusion is that while the continuum ratio is
close to the Casimir Scaling value of 1.5, there is a significant
and robust discrepancy at the $\sim 1.5\%$ level.

We have repeated the above calculation for SU(4), SU(6) and SU(8)
and the resulting continuum values, using $O(a^2)$ extrapolations,
are presented in Table~\ref{table_sigk}. The first error is
statistical and the second is intended to give some idea of
the maximum possible systematic error, by looking at what one
obtains from various combinations of $S$ and $D$ fits to
$\sigma_k$ and $\sigma_f$. (As such it is not an estimate of the 
actual systematic error which would undoubtedly be significantly 
smaller.) Some remarks. Firstly, for SU(6) our best fit has 
a rather poor, but not impossible,  $\chi^2$. We have therefore
doubled the errors so as to achieve a reasonable confidence level.
This will ensure that the $N=6$ value does not distort our
analysis of the $N$-dependence in the next Section.
Second remark: for $N=8$ one of the two finite volume calculations
has a very poor $\chi^2$ when we attempt to extract the string
tensions. We therefore do not include the resulting values, that have
very small errors but very small confidence levels, in our
continuum extrapolations, but instead
take from that study the string tension as calculated from the lattice
whose size (in physical units) is the same as we use at other
values of $\beta$. (Some freedom as to which values to include 
at those $\beta$ at which we perform finite volume
studies, as well as some freedom in the fitting process, explains
why the SU(5) values in Table~\ref{table_sigk} do not coincide
exactly with those in Table~\ref{table_n5cont}.) Note that as $k$ 
increases, the loop energy also increases, and the errors become less
reliable. Nonetheless, despite these caveats, we see from 
Table~\ref{table_sigk} that we can confidently conclude that while
the string tension ratios are remarkably close to the Casimir
Scaling values, and much further away from (MQCD) Sine Scaling, there
is a definite discrepancy that is typically at the $1-2\%$ level.

It is interesting to note that if we take the actual values of 
$\surd\sigma_f/g^2N$ as obtained in
\cite{gs_k1}
together with the values of $\sigma_k$ in eqn(\ref{eqn_kNair}),
we obtain the values of $\sigma_k/\sigma_f$ listed in the last 
column of Table~\ref{table_sigk}.
We see that these values are largely consistent with our 
calculated numbers. That is to say, while the prediction of
\cite{nair}
for $\sigma_f$ is too high by $\sim 1-2\%$, the prediction
for $\surd\sigma_k/g^2N$ with $k>1$ is actually consistent with the values
we obtain. This is particularly significant for the $k=2$
values which are the ones that we determine most accurately 
and reliably.

\section{$N$-dependence}
\label{section_N}

An interesting feature of the Casimir Scaling hypothesis in 
eqn(\ref{eqn_CS})
is that if we expand the expression for $\sigma_k/\sigma_f$ in powers
of $1/N$ at fixed $k$, we see that the leading correction is $O(1/N)$
rather than the $O(1/N^2)$ that one would expect from the usual
colour counting rules. This observation
\cite{mt_largeN02}
has generated some controversy
\cite{aamsN,ckahmN}.
It is therefore interesting to see if we can learn something about the 
power of the leading correction from a  more model-independent
analysis of our results.

\begin{figure}[htb]
\centerline{
\includegraphics[width=10cm]{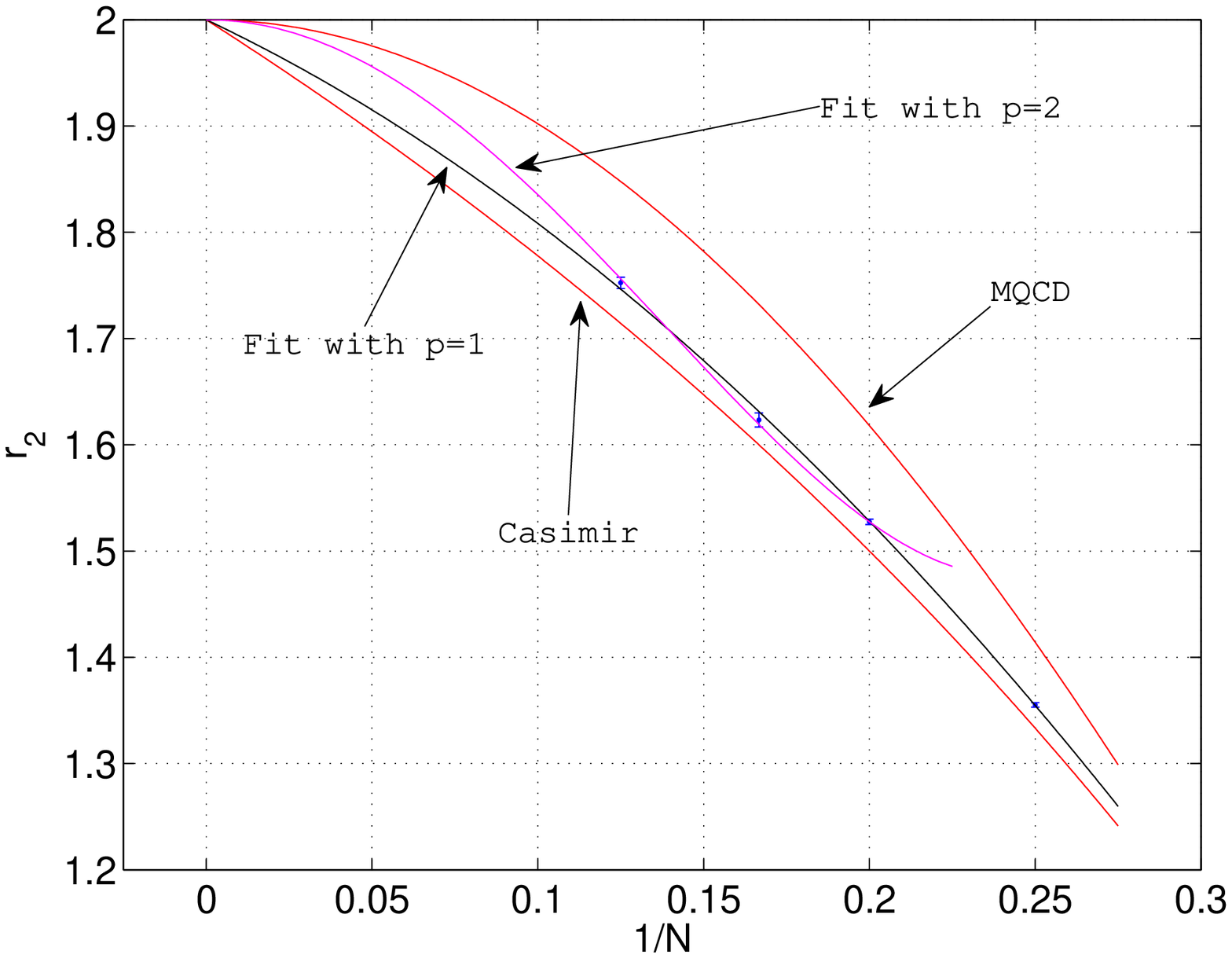}
}
\caption{Calculated values of $r_2 = \sigma_{k=2}/\sigma_f$ and
fits as discussed in Section~\ref{section_N}.}
\label{fig_kNa}
\end{figure}

We begin with our results for the $k=2$ continuum string tensions,
as these are more accurate than those for larger $k$. We attempt
a fit with the ansatz
\begin{equation}
\frac{\sigma_{k=2}}{\sigma_f}
=
2 - \frac{a}{N^p} - \frac{b}{N^{2p}}
\label{eqn_kNdep1} 
\end{equation}
using the values $p=1$ and $p=2$, and show the best such fits in
Fig.\ref{fig_kNa}.

For $p=1$ our best fit gives $a=1.48(5)$ and $b=4.42(21)$, with 
$\chi^2/n_{df}\simeq 2.7/2 = 1.35$. Thus we infer that an $O(1/N)$ correction 
is quite consistent
with our $k=2$ calculations even if, as we see in Fig.\ref{fig_kNa},
the full Casimir Scaling prediction is not. It is also interesting to
note that this best fit is consistent with the trivial value of 
${\sigma_{k=2}}/{\sigma_f}=1$ for SU(3), even though this value
has not been included in the fit.
For $p=2$, on the other hand, we find no acceptable fit at all, and in
order to proceed we have to drop the $N=4$ point from the fit. 
We now only have one degree of freedom, and although we obtain
an acceptable fit, with $a=18.0(5)$ and $b=-155(12)$, for this
reduced dataset, we see  
from Fig.\ref{fig_kNa} that it has a  wavy behaviour
that is characteristic of an inappropriate functional form
being forced onto data using large cancellations between different
terms. This perception is reinforced when we note how much this fit 
deviates from the SU(4) value.
We conclude from all this that our $k=2$ results strongly disfavour
a conventional $O(1/N^2)$ correction.

\begin{figure}[htb]
\centerline{
\includegraphics[width=10cm]{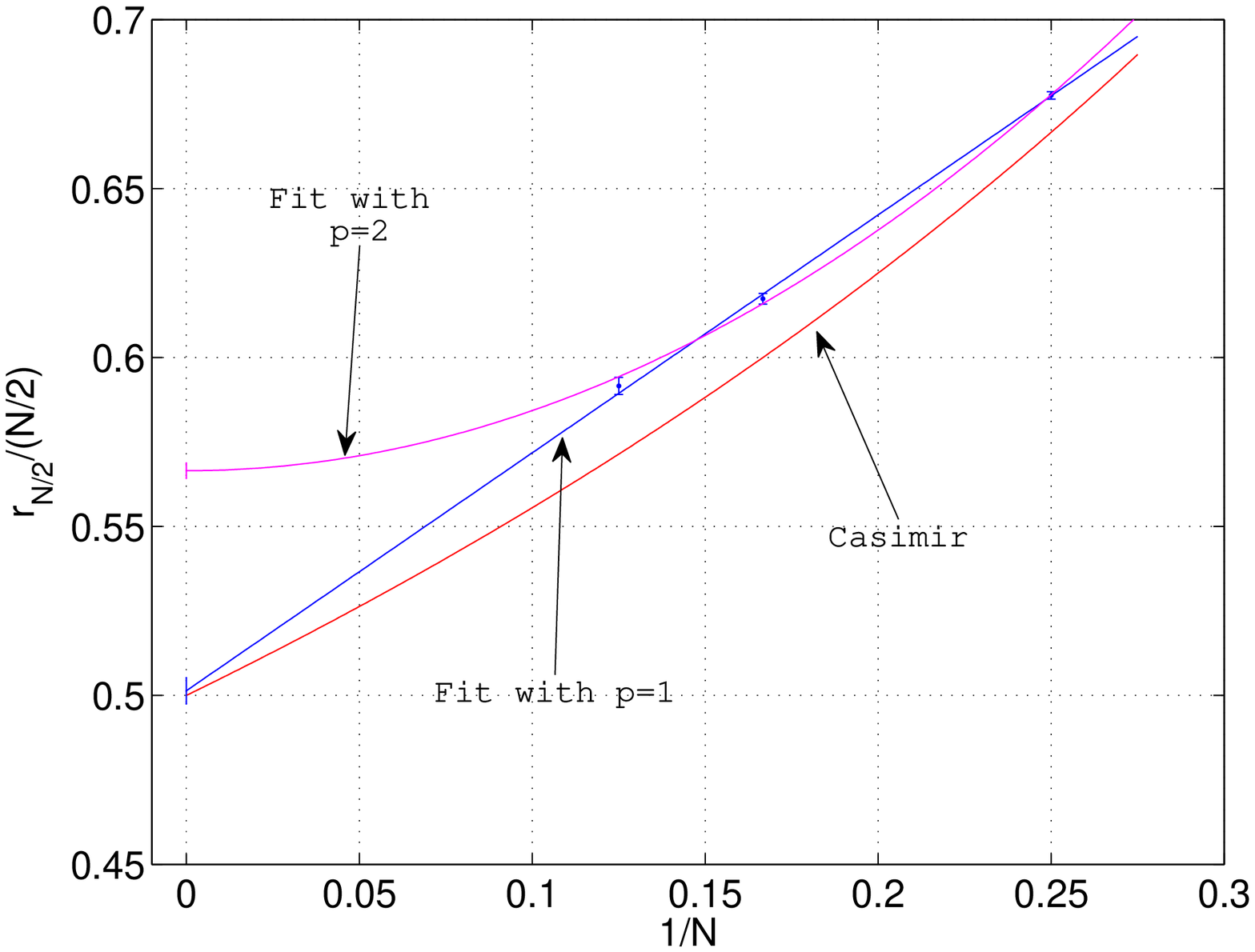}
}
\caption{Calculated values of $r_{N/2} = \sigma_{k=N/2}/\sigma_f$ and
fits as discussed in Section~\ref{section_N}.}
\label{fig_kNb}
\end{figure}

Complementary to the above large $N$ limit, in which $k$ is kept fixed, 
is the limit where $k/N$ is held fixed. The extreme example is
$k=N/2$; i.e. $k=2,3,4$ for $N=4,6,8$ respectively. Of course
the $k=4$ string is quite massive, so it will have a large
statistical error, and quite possibly a large systematic error
as well. Proceeding anyway, we fit with the ansatz
\begin{equation}
\frac{\sigma_{k=N/2}}{\sigma_f}
=
\frac{N}{2}\left\{
a + \frac{b}{N^p}
\right\}
\label{eqn_kNdep2} 
\end{equation}
with $p=1,2$. As we see in Fig.\ref{fig_kNb},
the best $p=1$ fit is acceptable (with $\chi^2 \simeq 1.5$ for
1 degree of freedom) and
gives $a=0.501(4)$ and $b=0.70(2)$. It is interesting to note 
that this value of $a$ is consistent with what one obtains 
at $N=\infty$ from 
Casimir scaling, although, as we see in Fig.\ref{fig_kNb},
the latter is actually far from the calculated ratios at finite $N$.
The $p=2$ best fit is  noticeably worse, with  $\chi^2/n_{df}\simeq
2.0$  and giving $a=0.567(3)$ and $b=1.78(5)$, but it is not so bad
that it can be completely excluded. Nonetheless it mildly reinforces 
our earlier conclusion that $O(1/N^2)$ corrections are disfavoured.

\section{Excited strings : group or centre?}
\label{section_group}

As we have seen (here and in earlier work) there do indeed exist 
non-trivial $k$-string bound
states. The ground state  $k$-string is stable, with string tension 
$\sigma_k < k\sigma_f$ at finite $N$, and we have provided strong
evidence that it falls into the universality class of the
simple Nambu-Goto string model. Thus one may conjecture that
there will be a spectrum of Nambu-Goto-like excitations
of this $k$-string, at long distances, for which the scale is 
set by  $\sigma_k$.

We also found that the string tension of the ground state 
$k$-string is remarkably close to the prediction of Casimir Scaling
i.e. that it is  proportional to the quadratic Casimir of the
totally antisymmetric representation of $k$ fundamentals. 
If this is more than mere coincidence, it implies that we should 
also expect to see some sign of flux tubes corresponding to other
representations, with string tensions (approximately) proportional to 
the corresponding Casimirs. Of course, since these Casimirs are larger,
such flux tubes will not be stable, but can be screened by gluons to the
totally antisymmetric one. If the amplitude for such screening is
large, such `resonant' states will be `broad', difficult to identify,
and will presumably play no significant dynamical role.
However if the amplitude for such screening is small enough, 
then it should be possible to identify such `resonant' flux tubes in 
the excited state spectrum of the $k$-strings. Indeed, in this case,
one may ask if such a `resonant' flux tube possesses its own spectrum
of Nambu-Goto-like excitations.

By contrast, in the free Nambu-Goto string model, which has 
been found to work very well for fundamental strings that wind once 
around a torus,
\cite{gs_k1,es_k1}
there are of course no bound state $k$-strings; we simply have 
$\sigma_k = k\sigma_f$ (as indeed one expects to have in the SU($N$)
gauge theory at $N=\infty$).
One may speculate that the observed binding can be encoded
through some correction terms to the Nambu-Goto action, leading 
to a bound state string for each value of $k$. Moreover it 
is possible that each of these stable bound states
will be accompanied by its own tower of Nambu-Goto-like excited
states, once the string is long enough. However, while there
may be additional `resonant' string excitations, it is not clear how
these correction terms could embody the detailed group structure of
the SU($N$) gauge theory in any natural fashion, and so one would not 
expect the excited states to be related in any way to the different 
representations of the SU($N$) gauge group. 

There is some evidence from calculations of potentials between sources
in different representations that the corresponding fluxes do form
flux tubes
\cite{BaliDeldar}, 
but such calculations do not go to large enough separations to
be unambiguous. There is also some evidence from earlier calculations
of closed loops, that the lowest excited 
$k=2$ strings fall approximately into symmetric and anti-symmetric 
representations 
\cite{old_k_oxford},
although the observed near-degeneracies in these states
\cite{blmtuw-glue04}
have been interpreted as saying that these are in fact the same states,
and that the categorisation at a given $k$ into different representions
is illusory: that is to say, confining strings only know about the center 
of the group
\cite{aabl}.

To determine what is actually the case, one needs accurate
calculations of a significant number of excited states at each $k$.
This requires a large basis of operators enabling one to obtain good
overlaps onto these excited states, paralleling the $k=1$ calculations in 
\cite{gs_k1}.
Such a calculation is under way
\cite{es_k}.
In the meantime, it is interesting to see what we can extract
from the present calculation which has a rather limited basis, with 
overlaps on to the excited $k$-string states that are only moderately
good.

We restrict ourselves here to the $k=2$ excited state spectrum which
is the one that we can determine most accurately. The operators whose 
correlators we have calculated are of the form $\mathrm{Tr} l^2$ and  
$\{\mathrm{Tr} l\}^2$ where $l$ is a Polyakov loop. There will be 
such operators corresponding to each of the smearing/blocking levels
and this provides us with a non-trivial  basis for our variational 
calculation of the $k=2$ spectrum.
This basis of operators allows us to construct Polyakov loops
in the totally symmetric (2S) and antisymmetric (2A) representations:
\begin{equation}
\mathrm{Tr}_{2A}l
=
\mathrm{Tr}l^2
-
\{\mathrm{Tr} l\}^2 \quad;\quad
\mathrm{Tr}_{2S}l
=
\mathrm{Tr}l^2
+
\{\mathrm{Tr} l\}^2.
\label{eqn_lk2AS} 
\end{equation}
To construct $k=2$ operators in other representations would require
higher powers, e.g. $\mathrm{Tr}l^3 \mathrm{Tr}l^\dagger$, which we
have not calculated here. Thus we shall limit our analysis to
the $k=2A$ and $k=2S$ representations. Using the basic operators
in eqn(\ref{eqn_lk2AS}), we construct $\vec{p}=0$ operators at time
$t$ by summing up $\mathrm{Tr}_{2A,2S}l$ over the spatial
coordinates. We call the resulting operators $\Phi_{2A}(t)$ and
$\Phi_{2S}(t)$ where there is an additional suppressed label, $b$,
for the blocking level of the gauge links used in the construction
of $l$. These are our basis operators and our variational procedure
will (ideally) produce the linear combinations that are closest
to the actual energy eigenstates. 

One can immediately learn something interesting from these basis operators.
Consider the normalised overlap of a symmetric operator $\Phi_{2S}$
at blocking levels $b_S$ with an antisymmetric operator $\Phi_{2A}$ 
at blocking levels $b_A$:
\begin{equation}
O_{AS}(b_A,b_S)
=
\frac{\langle \Phi^\dagger_{2A}(0) \Phi_{2S}(0) \rangle}
{\langle \Phi^\dagger_{2A}(0) \Phi_{2A}(0) \rangle^{1/2}
\langle \Phi^\dagger_{2S}(0) \Phi_{2S}(0) \rangle^{1/2}}.
\label{eqn_OAS} 
\end{equation}
If these overlaps turn out to be large, then the screening/mixing
effects are large, and there is not much point in trying to find
different string states corresponding to different representations. 
If they are small, then it becomes interesting to determine the 
representation content of the ground and excited $k$-string states.

In Table~\ref{table_OAS} we list the values of these overlaps as 
calculated on a $24^2 32$ lattice in SU(5), at a value of $\beta$ 
that corresponds to $a\surd\sigma_f \simeq 0.13$. This is both 
close to the continuum limit and ensures that the string is long
enough, $l\surd\sigma_f \simeq 3.1$. Our results show that the
overlaps are remarkably small. Indeed they are all consistent
with zero, within very small errors, except for those that involve 
the very largest blocking level. Blocked links at this level
spread right across the spatial volume and we have convincing 
evidence from calculations on larger volumes that
the somewhat enhanced overlap is primarily a finite volume effect.
Nonetheless even these non-zero values are in fact very small.
This provides striking evidence that the screening dynamics
is a weak perturbation on the classification of states into
representations of SU($N$).

\begin{figure}[htb]
\centerline{
\includegraphics[width=10cm]{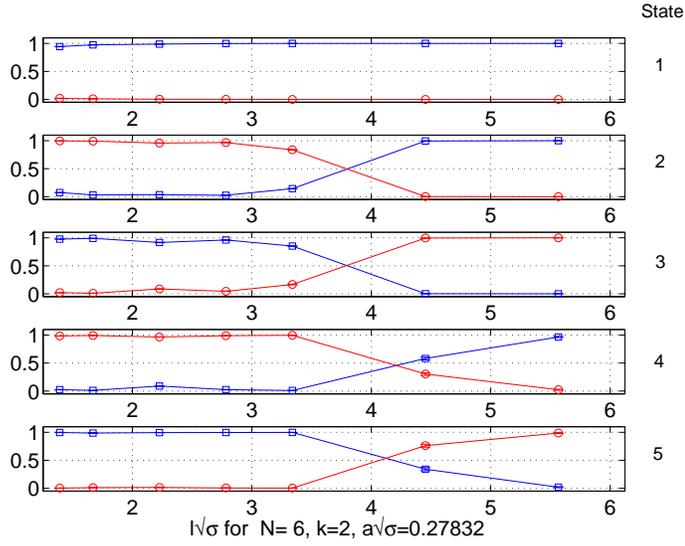}
}
\caption{ Overlaps of the five lowest 
states in the $k=2$ spectrum of $SU(6)$ at $\beta=59.40$ ($a\simeq 
0.12$ fm) vs. the string length. In blue(red) are the overlaps 
onto the antisymmetric(symmetric) representation. }
\label{fig_k2over}
\end{figure}

We now turn to our variational estimates of the the $k=2$ string 
eigenstates, $\Psi_J$, that we obtain
using the full basis of $\Phi_{2A}(t)$ and $\Phi_{2S}(t)$ 
operators. We project the resulting eigenstates onto the $k=2A$ and  
$k=2S$ subspaces. (Which we do by forming  an orthogonal 
basis out of our non-orthogonal $\Phi_{2A,2S}$ operators.)
We call the corresponding overlaps $O_{JA}$ and  $O_{JS}$,
where  we expect $|O_{JA}|^2+|O_{JS}|^2 \simeq 1$ since
the  $k=2A$ and  $k=2S$ subspaces are nearly orthogonal. 
If the subspaces were exactly
orthogonal then it would of course immediately follow that the
eigenstates would fall into exact $k=2A$ and $k=2S$ representations.
What we find is that they appear to do so to quite 
a good approximation. An example is shown in Fig.\ref{fig_k2over}.
Here we plot the overlaps $|O_{JA}|^2, |O_{JS}|^2$ for the
first five $k=2$ states, $J=1,...,5$, as a function of the string
length $l$, in a calculation in SU(6) at $\beta=59.4$ (which
corresponds to $a\surd\sigma_f \simeq 0.28$). What we see
is that the states are almost entirely either pure $k=2A$ or
$k=2S$ except at certain values of $l$ where the state
changes from one representation to the other.
Since this always happens to pairs of states at the same value
of $l$ and involves an opposite change in representation
in this pair, it is clear what is happening: as we increase $l$ 
the two energy levels cross at that $l$. So, for example, in 
Fig.\ref{fig_k2over} the states that are $\Psi_2$ and $\Psi_3$ for 
$l\surd\sigma_f < 3$ exchange their ordering in energy 
for $l\surd\sigma_f > 4$. And a similar exchange occurs
for $\Psi_4$ and $\Psi_5$, but at a larger value of $l$.
What we see, therefore, is that the $k$-string states belong
to particular SU($N$) representations to a very good approximation.
The apparent near-degenaracies observed in
\cite{blmtuw-glue04}
are in fact accidental: they arise from the (natural) fact
that the various excited states have energies that vary differently 
with $l$, so that they cross as $l$ increases and when they do so
we have degeneracies. 
As we have just seen, the first $k=2$ crossing occurs at
$l\surd\sigma_f \sim 3.5$ which is precisely where such
degeneracies have been previously noted to occur 
\cite{old_k_oxford,blmtuw-glue04}.
It is unambiguous that what we are seeing here is not a single state,
as suggested in
\cite{aabl},
but rather two that are nearly degenerate at appropriate values of $l$.

\begin{figure}[htb]
\centerline{
\includegraphics[width=10cm]{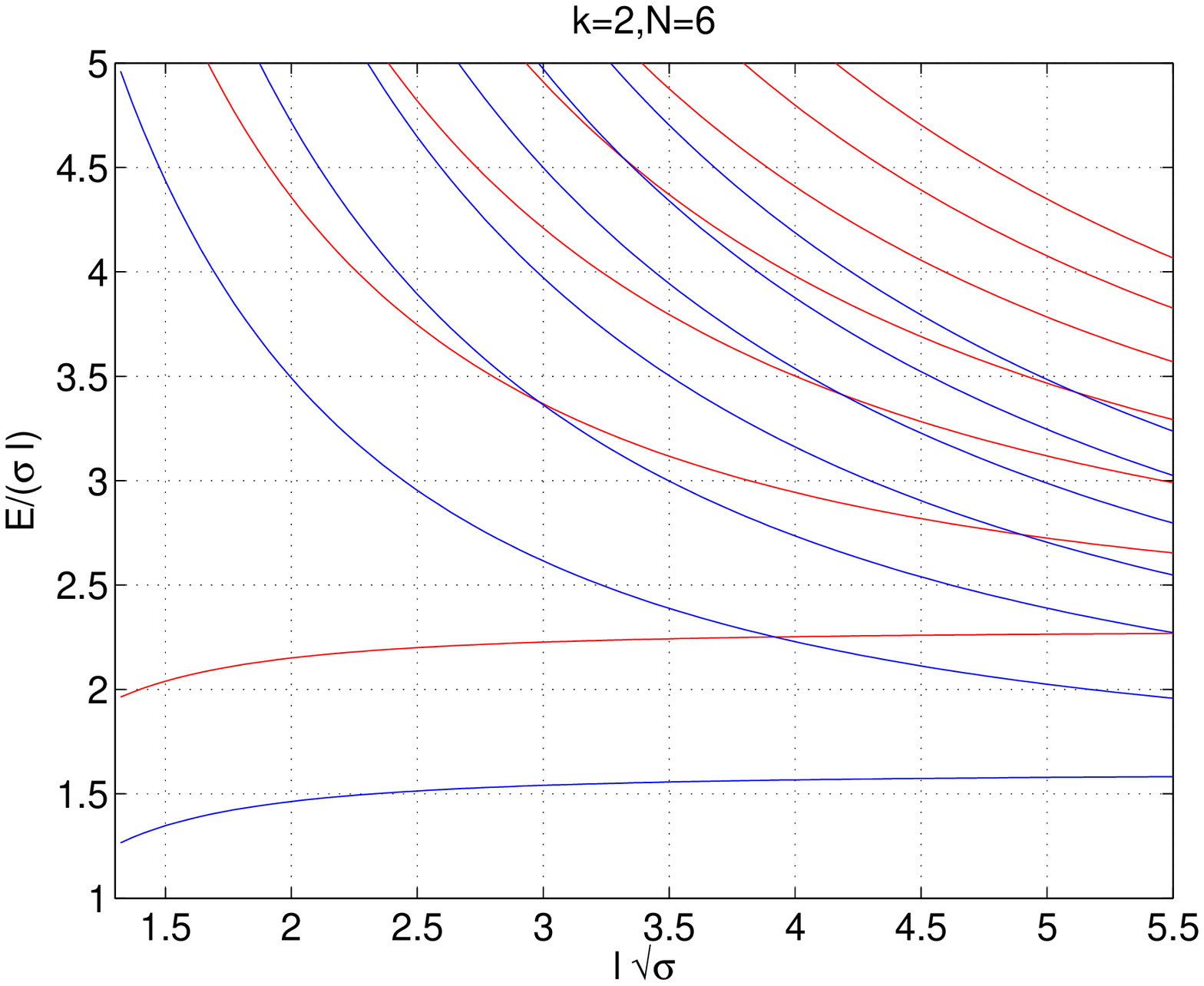}
}
\caption{ The spectrum of energies vs. the string length
from the Nambu-Goto model for $SU(6)$ and $k=2$. 
Blue(red) denotes the energies of the 
antisymmetric(symmetric) representation.} 
\label{fig_k2spectrumNG}
\end{figure}

If we focus on the lightest three states in Fig.\ref{fig_k2over},
we see that the ground state is always totally anti-symmetric,
while the first excited state is initially symmetric and then,
for $l\surd\sigma_f > 4$, antisymmetric. This very different
dependence on $l$ between the ground state $k=2S$ state and
the first excited  $k=2A$ state is easily understood if we
think of there being two separate Nambu-Goto towers of states
labelled by the $k=2A$ and $k=2S$ representations respectively. 
In the Nambu-Goto model the energy levels would be given by
\begin{equation}
E_n(l) 
=
\sigma_{\cal{R}} l 
\left( 
1 + 8\pi \frac{\pi}{\sigma_{\cal{R}} l^2} 
\left(n -\frac{D-2}{24}\right)
\right)^\frac{1}{2}
\label{eqn_NGe} 
\end{equation}
where $\sigma_{\cal{R}}$ is the string tension of the flux in
representation $\cal{R}$ and $n$ is an integer that
counts the `phonon' excitations along the string. For the
ground string state we have $n = 0$ and $E_0(l)$
increases approximately linearly with $l$ in the range of
$l$ relevant here. On the other hand the first excited
state with $n =1 $ will clearly have a very different 
variation with $l$. This is illustrated in 
Fig.\ref{fig_k2spectrumNG} where we plot the Nambu-Goto
energy levels (divided by $\sigma_f l$) for towers of states in 
the $k=2A$ and  $k=2S$ representations of SU(6) using the string 
tensions that are predicted by Casimir Scaling.
One sees how the very different $l$-dependence
of the ground state $k=2S$ string and the first excited $k=2A$ 
string ensures that they cross for a value of $l\surd\sigma_f$
close to where we observe the crossing in Fig.\ref{fig_k2over}.
Our estimates of the actual energy levels are very roughly
consistent with such a scenerio, although a reliable and precise
comparison must await a dedicated study of the $k>1$ energy spectrum
\cite{es_k}.
In the meanwhile what we have is very good evidence that
the $k$-string states fall, to a good approximation, into
representations of SU($N$), and strong indications, from the crossings
of the states, that there exists a ground state flux tube in
representations such as $k=2S$ where it is not completely stable.

\section{Conclusions}
\label{section_conc}

In this paper we have studied $k$-strings that wind around a spatial 
torus of length $l$. This setup is convenient because it involves no 
sources so one knows that, for $l\geq 1/T_c$, all that one has is 
a closed string-like flux tube. It provides a clean way to determine the 
effective string theory describing the flux tube. Once this question
has been addressed the second, equally important question of how the 
effective string theory at long distances matches onto short-distance
asymptotic freedom, can be conveniently addressed through the
calculation of the potential between sources as a function of the
distance between these sources. For example, one can ask
whether the running coupling in the `potential scheme'
\begin{equation}
V(r) 
\equiv
-C \frac{\alpha_v(r)}{r}
\label{eqn_alphav} 
\end{equation}
($C$ is the quadratic Casimir for the representation of the
sources) develops a non-analyticity at some critical distance 
as $N\to\infty$.

The main conclusions of our study of such closed $k$-strings 
are as follows.

$\bullet$ We find that just like the fundamental $k=1$ flux tube, $k = 2$
flux tubes are described by an effective string theory 
that is in the universality class of the Nambu-Goto model.
If one parameterises the correction to Nambu-Goto by an appropriate
power of $1/\sigma l^2$, one finds that the coefficient is $O(1)$;
that is to say, it is neither small nor large. This is in contrast
to the case of the fundamental string, where such corrections are
remarkably small
\cite{gs_k1}.

$\bullet$ The $k\geq 3$ flux tubes are also consistent
with being in the simple bosonic string universality class.
While the statistical errors are larger than for $k=2$,
they also exclude a central charge $\propto \sigma_k/\sigma_f$
as suggested in
\cite{gliozzi}.

$\bullet$ The ratio of the ground state string tensions,
$\sigma_k/\sigma_f$, is within $1-2\%$ of Casimir Scaling, but this 
discrepancy is statistically significant, and it is robust against 
systematic errors. On the other hand, the values of $\sigma_k/g^2N$
are consistent with the predictions of
\cite{nair}
for $k>1$; in particular for $k=2$ where we have very precise
results. From this point of view, the small breakdown of Casimir 
Scaling that we observe can be entirely attributed to the fact that 
the predicted value of $\sigma_f$ in
\cite{nair}
is slightly higher than its actual value as obtained in 
\cite{gs_k1}.
One should perhaps not read too much significance into this
agreement. The important point is that any discrepancy between
the predictions for  $\sigma_k$ in
\cite{nair}
and the true values are no larger for $k>1$ strings than 
they are for the fundamental string.

$\bullet$ An analysis of how the $k=2$ string tension varies 
with $N$ strongly suggests that the corrections to 
the $N=\infty$ limit, $\sigma_{k=2} = 2\sigma_f$, come in
powers of $1/N$ (as in Casimir Scaling) rather than in
powers of $1/N^2$, as suggested by standard colour counting
arguments. There is also some statistically weaker evidence 
for this from our analysis of the $N$-dependence of $\sigma_{k=N/2}$
for $N=4,6,8$.

$\bullet$ The $k$-string eigenstates fall, very accurately, into 
representations of SU($N$). That is to say, $k$-strings know about 
the full group representation of the flux, and not just about its 
transformation properties under the centre. The near-degeneracies 
previously observed between some states assigned to different 
representations, should be interpreted as level crossing (as a 
function of the string length $l$). The intriguing possibility
that stable $k$-strings (and perhaps even unstable `resonant' 
strings) may possess their own towers of Nambu-Goto states, 
receives some support from preliminary precision results on 
the $k$-string spectrum
\cite{es_k}.

\section*{Acknowledgements}

We acknowledge useful discussions with Ofer Aharony,
Nick Dorey, James Drummond,  Pietro Giudice, Jeff Greensite,
Dimitra Karabali, V. P. Nair, David Tong, Peter Weisz
and, in particular, with Andreas Athenodorou, with whom
we are collaborating on a closely related project.
The computations were performed on resources funded
by Oxford and EPSRC.
BB was supported in part by the U.S. Department of Energy under 
Grant No. DE-FG02-96ER40956.

\vfill\eject

\begin{table}
\begin{center}
\begin{tabular}{|c|c|c||c|c|c|}\hline
\multicolumn{3}{|c|}{SU(4) $\beta=32.0$ } &
\multicolumn{3}{|c|}{SU(5) $\beta=80.0$ }  \\ \hline 
$L_x$ &  $aE_0(k=1)$  &  $aE_0(k=2)$ &
$L_x$ &  $aE_0(k=1)$  &  $aE_0(k=2)$  \\ \hline
6   & 0.1637(8)  & 0.2428(17) & 10  & 0.1015(6)  &  0.1720(20) \\
8   & 0.2976(6)  & 0.4241(11) & 12  & 0.1519(5)  &  0.2485(12) \\
10  & 0.4075(5)  & 0.5716(10) & 16  & 0.2344(5)  &  0.3747(9)  \\
12  & 0.5105(7)  & 0.7043(23) & 20  & 0.3089(6)  &  0.4867(11) \\
16  & 0.7073(13) & 0.9656(35) & 24  & 0.3827(6)  &  0.5938(15) \\
    &            &            & 32  & 0.5242(11) &  0.8090(28) \\ \hline
\end{tabular}
\caption{\label{table_V}
The masses of the lightest $k=1$ and $k=2$ flux loops as a function
of their length $l=aL_x$, for SU(4) and SU(5) at the indicated
values of $\beta=2N/g^2$.}
\end{center}
\end{table}

\begin{table}
\begin{center}
\begin{tabular}{|c|cc|cc|}\hline
\multicolumn{5}{|c|}{$\lim_{a\to 0}\sigma_{k=2}/\sigma_f$ ; SU(5) } \\ \hline
fit &  S  &  $\chi^2/n_{df}$ & D & $\chi^2/n_{df}$ \\ \hline
$O(a^2)$  & 1.528(4) & 1.3 & 1.524(7)  & 0.9 \\
$O(a^4)$  & 1.520(5) & 1.1 & 1.509(11) & 0.7 \\ \hline
\end{tabular}
\caption{\label{table_n5cont}
Continuum limits of the extrapolations in 
Figs\ref{fig_n5Scont},\ref{fig_n5Dcont}
with quality of best fit.}
\end{center}
\end{table}

\begin{table}
\begin{center}
\begin{tabular}{|c|c|c||c|c||c|}\hline
$k$ &  $N$ &  $\sigma_k/\sigma_f$  &
Casimir &  Sine  & Nair  \\ \hline
2 & 4 & \rule[-1mm]{0mm}{6mm}1.3552(22)${91\choose 29}$  & 1.333.. & 1.414..  & 1.361 \\
2 & 5 & \rule[-1mm]{0mm}{6mm}1.5275(26)${96\choose 0}$ & 1.5     & 1.618..  & 1.529 \\
2 & 6 & \rule[-1mm]{0mm}{6mm}1.6234(66)${98\choose 0}$  & 1.6     & 1.732..  & 1.629 \\
2 & 8 & \rule[-3mm]{0mm}{8mm}1.7524(51)${107\choose 30}$ & 1.714.. & 1.848..  & 1.741 \\ \hline
3 & 6 & \rule[-1mm]{0mm}{6mm}1.8522(48)${134\choose 0}$  & 1.8     & 2.0      & 1.832 \\
3 & 8 & \rule[-3mm]{0mm}{8mm}2.174(19)${14\choose 9}$   & 2.143.. & 2.414..  & 2.177 \\ \hline
4 & 8 & \rule[-3mm]{0mm}{8mm}2.366(11)${11\choose 6}$   & 2.286.. & 2.613..  & 2.322 \\ \hline
\end{tabular}
\caption{\label{table_sigk}
$k$-string tensions compared to Casimir and Sine Scaling conjectures
in eqns(\ref{eqn_CS},\ref{eqn_MQCD})
and the Nair prediction in eqn(\ref{eqn_kNair}). First error
statistical, second estimate of systematics.}
\end{center}
\end{table}

\begin{table}
\begin{center}
\begin{tabular}{|c|c|c||c|c|c||c|c|c|}\hline
$b_A$ &  $b_S$  &  $O_{AS}$ &
$b_A$ &  $b_S$  &  $O_{AS}$ &
$b_A$ &  $b_S$  &  $O_{AS}$ \\ \hline 
1 & 1 & -0.0003(8) & 2 & 1 & 0.0006(6)  & 3 & 1 & 0.0006(5) \\
1 & 2 & -0.0005(8) & 2 & 2 & 0.0005(7)  & 3 & 2 & 0.0005(7) \\
1 & 3 & -0.0006(8) & 2 & 3 & 0.0002(8)  & 3 & 3 & -0.0001(10) \\
1 & 4 &  0.0002(8) & 2 & 4 & 0.0006(10) & 3 & 4 & 0.0008(12) \\
1 & 5 &  0.0025(6) & 2 & 5 & 0.0047(8)  & 3 & 5 & 0.0066(11) \\ \hline
4 & 1 &  0.0006(6)  & 5 & 1 & -0.0002(5)  & & & \\
4 & 2 &  0.0005(9)  & 5 & 2 &  0.0005(8)  & & & \\
4 & 3 & -0.0003(11) & 5 & 3 &  0.0009(9)  & & & \\
4 & 4 &  0.0010(13) & 5 & 4 &  0.0038(12) & & & \\
4 & 5 &  0.0110(13) & 5 & 5 &  0.0290(15) & & &  \\ \hline
\end{tabular}
\caption{\label{table_OAS}
Overlaps of Polyakov loops in the k=2A and k=2S representations,
with blocking levels $b_A$ and $b_S$ respectively. In SU(5)
at $\beta=80.0$ on a $24^2 32$ lattice.}
\end{center}
\end{table}

\clearpage

\begin	{figure}[p]
\begin	{center}
\leavevmode
\setlength{\unitlength}{0.240900pt}
\ifx\plotpoint\undefined\newsavebox{\plotpoint}\fi
\sbox{\plotpoint}{\rule[-0.200pt]{0.400pt}{0.400pt}}%
\begin{picture}(1800,900)(0,0)
\font\gnuplot=cmr10 at 12pt
\gnuplot
\sbox{\plotpoint}{\rule[-0.200pt]{0.400pt}{0.400pt}}%
\put(350.0,150.0){\rule[-0.200pt]{4.818pt}{0.400pt}}
\put(325,150){\makebox(0,0)[r]{\ \ {$0$}}}
\put(1705.0,150.0){\rule[-0.200pt]{4.818pt}{0.400pt}}
\put(350.0,325.0){\rule[-0.200pt]{4.818pt}{0.400pt}}
\put(325,325){\makebox(0,0)[r]{\ \ {$0.5$}}}
\put(1705.0,325.0){\rule[-0.200pt]{4.818pt}{0.400pt}}
\put(350.0,500.0){\rule[-0.200pt]{4.818pt}{0.400pt}}
\put(325,500){\makebox(0,0)[r]{\ \ {$1$}}}
\put(1705.0,500.0){\rule[-0.200pt]{4.818pt}{0.400pt}}
\put(350.0,675.0){\rule[-0.200pt]{4.818pt}{0.400pt}}
\put(325,675){\makebox(0,0)[r]{\ \ {$1.5$}}}
\put(1705.0,675.0){\rule[-0.200pt]{4.818pt}{0.400pt}}
\put(350.0,850.0){\rule[-0.200pt]{4.818pt}{0.400pt}}
\put(325,850){\makebox(0,0)[r]{\ \ {$2$}}}
\put(1705.0,850.0){\rule[-0.200pt]{4.818pt}{0.400pt}}
\put(350.0,150.0){\rule[-0.200pt]{0.400pt}{4.818pt}}
\put(350,100){\makebox(0,0){\ {$1$}}}
\put(350.0,830.0){\rule[-0.200pt]{0.400pt}{4.818pt}}
\put(694.0,150.0){\rule[-0.200pt]{0.400pt}{4.818pt}}
\put(694,100){\makebox(0,0){\ {$2$}}}
\put(694.0,830.0){\rule[-0.200pt]{0.400pt}{4.818pt}}
\put(1038.0,150.0){\rule[-0.200pt]{0.400pt}{4.818pt}}
\put(1038,100){\makebox(0,0){\ {$3$}}}
\put(1038.0,830.0){\rule[-0.200pt]{0.400pt}{4.818pt}}
\put(1381.0,150.0){\rule[-0.200pt]{0.400pt}{4.818pt}}
\put(1381,100){\makebox(0,0){\ {$4$}}}
\put(1381.0,830.0){\rule[-0.200pt]{0.400pt}{4.818pt}}
\put(1725.0,150.0){\rule[-0.200pt]{0.400pt}{4.818pt}}
\put(1725,100){\makebox(0,0){\ {$5$}}}
\put(1725.0,830.0){\rule[-0.200pt]{0.400pt}{4.818pt}}
\put(350.0,150.0){\rule[-0.200pt]{331.237pt}{0.400pt}}
\put(1725.0,150.0){\rule[-0.200pt]{0.400pt}{168.630pt}}
\put(350.0,850.0){\rule[-0.200pt]{331.237pt}{0.400pt}}
\put(150,700){\makebox(0,0){\Large{$c_{eff}$}}}
\put(1012,25){\makebox(0,0){\large{$l\surd\sigma_k$}}}
\put(350.0,150.0){\rule[-0.200pt]{0.400pt}{168.630pt}}
\put(524.0,520.0){\rule[-0.200pt]{0.400pt}{2.168pt}}
\put(514.0,520.0){\rule[-0.200pt]{4.818pt}{0.400pt}}
\put(514.0,529.0){\rule[-0.200pt]{4.818pt}{0.400pt}}
\put(672.0,497.0){\rule[-0.200pt]{0.400pt}{3.613pt}}
\put(662.0,497.0){\rule[-0.200pt]{4.818pt}{0.400pt}}
\put(662.0,512.0){\rule[-0.200pt]{4.818pt}{0.400pt}}
\put(820.0,488.0){\rule[-0.200pt]{0.400pt}{6.504pt}}
\put(810.0,488.0){\rule[-0.200pt]{4.818pt}{0.400pt}}
\put(810.0,515.0){\rule[-0.200pt]{4.818pt}{0.400pt}}
\put(1042.0,474.0){\rule[-0.200pt]{0.400pt}{9.154pt}}
\put(1032.0,474.0){\rule[-0.200pt]{4.818pt}{0.400pt}}
\put(1032.0,512.0){\rule[-0.200pt]{4.818pt}{0.400pt}}
\put(450.0,525.0){\rule[-0.200pt]{35.653pt}{0.400pt}}
\put(450.0,515.0){\rule[-0.200pt]{0.400pt}{4.818pt}}
\put(598.0,515.0){\rule[-0.200pt]{0.400pt}{4.818pt}}
\put(599.0,505.0){\rule[-0.200pt]{35.412pt}{0.400pt}}
\put(599.0,495.0){\rule[-0.200pt]{0.400pt}{4.818pt}}
\put(746.0,495.0){\rule[-0.200pt]{0.400pt}{4.818pt}}
\put(746.0,502.0){\rule[-0.200pt]{35.653pt}{0.400pt}}
\put(746.0,492.0){\rule[-0.200pt]{0.400pt}{4.818pt}}
\put(894.0,492.0){\rule[-0.200pt]{0.400pt}{4.818pt}}
\put(894.0,493.0){\rule[-0.200pt]{71.306pt}{0.400pt}}
\put(894.0,483.0){\rule[-0.200pt]{0.400pt}{4.818pt}}
\put(524,525){\circle*{24}}
\put(672,505){\circle*{24}}
\put(820,502){\circle*{24}}
\put(1042,493){\circle*{24}}
\put(1190.0,483.0){\rule[-0.200pt]{0.400pt}{4.818pt}}
\put(607.0,632.0){\rule[-0.200pt]{0.400pt}{4.095pt}}
\put(597.0,632.0){\rule[-0.200pt]{4.818pt}{0.400pt}}
\put(597.0,649.0){\rule[-0.200pt]{4.818pt}{0.400pt}}
\put(779.0,561.0){\rule[-0.200pt]{0.400pt}{6.986pt}}
\put(769.0,561.0){\rule[-0.200pt]{4.818pt}{0.400pt}}
\put(769.0,590.0){\rule[-0.200pt]{4.818pt}{0.400pt}}
\put(951.0,420.0){\rule[-0.200pt]{0.400pt}{20.236pt}}
\put(941.0,420.0){\rule[-0.200pt]{4.818pt}{0.400pt}}
\put(941.0,504.0){\rule[-0.200pt]{4.818pt}{0.400pt}}
\put(1209.0,441.0){\rule[-0.200pt]{0.400pt}{26.981pt}}
\put(1199.0,441.0){\rule[-0.200pt]{4.818pt}{0.400pt}}
\put(1199.0,553.0){\rule[-0.200pt]{4.818pt}{0.400pt}}
\put(522.0,641.0){\rule[-0.200pt]{41.194pt}{0.400pt}}
\put(522.0,631.0){\rule[-0.200pt]{0.400pt}{4.818pt}}
\put(693.0,631.0){\rule[-0.200pt]{0.400pt}{4.818pt}}
\put(693.0,575.0){\rule[-0.200pt]{41.435pt}{0.400pt}}
\put(693.0,565.0){\rule[-0.200pt]{0.400pt}{4.818pt}}
\put(865.0,565.0){\rule[-0.200pt]{0.400pt}{4.818pt}}
\put(865.0,462.0){\rule[-0.200pt]{41.435pt}{0.400pt}}
\put(865.0,452.0){\rule[-0.200pt]{0.400pt}{4.818pt}}
\put(1037.0,452.0){\rule[-0.200pt]{0.400pt}{4.818pt}}
\put(1037.0,497.0){\rule[-0.200pt]{82.870pt}{0.400pt}}
\put(1037.0,487.0){\rule[-0.200pt]{0.400pt}{4.818pt}}
\put(607,641){\circle{24}}
\put(779,575){\circle{24}}
\put(951,462){\circle{24}}
\put(1209,497){\circle{24}}
\put(1381.0,487.0){\rule[-0.200pt]{0.400pt}{4.818pt}}
\put(350,500){\usebox{\plotpoint}}
\put(350.0,500.0){\rule[-0.200pt]{331.237pt}{0.400pt}}
\end{picture}

\end	{center}
\caption{The effective coefficient of the $\pi/3\sigma l^2$ 
term in the Nambu-Goto expression for the ground state energy,
eqn(\ref{eqn_NGceff}), for the range of lengths indicated. For $k=1$, 
$\bullet$, and $k=2$, $\circ$, strings in SU(4) at $\beta=32.0$.}
\label{fig_NGceffN4}
\end 	{figure}
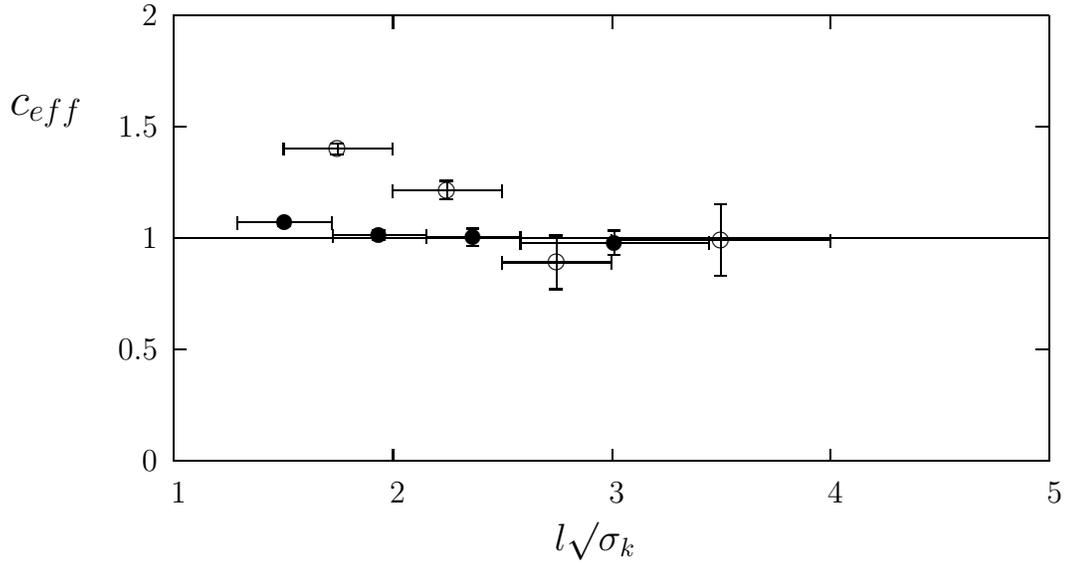

\begin	{figure}[p]
\begin	{center}
\leavevmode
\setlength{\unitlength}{0.240900pt}
\ifx\plotpoint\undefined\newsavebox{\plotpoint}\fi
\sbox{\plotpoint}{\rule[-0.200pt]{0.400pt}{0.400pt}}%
\begin{picture}(1800,900)(0,0)
\font\gnuplot=cmr10 at 12pt
\gnuplot
\sbox{\plotpoint}{\rule[-0.200pt]{0.400pt}{0.400pt}}%
\put(350.0,150.0){\rule[-0.200pt]{4.818pt}{0.400pt}}
\put(325,150){\makebox(0,0)[r]{\ \ {$0$}}}
\put(1705.0,150.0){\rule[-0.200pt]{4.818pt}{0.400pt}}
\put(350.0,325.0){\rule[-0.200pt]{4.818pt}{0.400pt}}
\put(325,325){\makebox(0,0)[r]{\ \ {$0.5$}}}
\put(1705.0,325.0){\rule[-0.200pt]{4.818pt}{0.400pt}}
\put(350.0,500.0){\rule[-0.200pt]{4.818pt}{0.400pt}}
\put(325,500){\makebox(0,0)[r]{\ \ {$1$}}}
\put(1705.0,500.0){\rule[-0.200pt]{4.818pt}{0.400pt}}
\put(350.0,675.0){\rule[-0.200pt]{4.818pt}{0.400pt}}
\put(325,675){\makebox(0,0)[r]{\ \ {$1.5$}}}
\put(1705.0,675.0){\rule[-0.200pt]{4.818pt}{0.400pt}}
\put(350.0,850.0){\rule[-0.200pt]{4.818pt}{0.400pt}}
\put(325,850){\makebox(0,0)[r]{\ \ {$2$}}}
\put(1705.0,850.0){\rule[-0.200pt]{4.818pt}{0.400pt}}
\put(350.0,150.0){\rule[-0.200pt]{0.400pt}{4.818pt}}
\put(350,100){\makebox(0,0){\ {$1$}}}
\put(350.0,830.0){\rule[-0.200pt]{0.400pt}{4.818pt}}
\put(694.0,150.0){\rule[-0.200pt]{0.400pt}{4.818pt}}
\put(694,100){\makebox(0,0){\ {$2$}}}
\put(694.0,830.0){\rule[-0.200pt]{0.400pt}{4.818pt}}
\put(1038.0,150.0){\rule[-0.200pt]{0.400pt}{4.818pt}}
\put(1038,100){\makebox(0,0){\ {$3$}}}
\put(1038.0,830.0){\rule[-0.200pt]{0.400pt}{4.818pt}}
\put(1381.0,150.0){\rule[-0.200pt]{0.400pt}{4.818pt}}
\put(1381,100){\makebox(0,0){\ {$4$}}}
\put(1381.0,830.0){\rule[-0.200pt]{0.400pt}{4.818pt}}
\put(1725.0,150.0){\rule[-0.200pt]{0.400pt}{4.818pt}}
\put(1725,100){\makebox(0,0){\ {$5$}}}
\put(1725.0,830.0){\rule[-0.200pt]{0.400pt}{4.818pt}}
\put(350.0,150.0){\rule[-0.200pt]{331.237pt}{0.400pt}}
\put(1725.0,150.0){\rule[-0.200pt]{0.400pt}{168.630pt}}
\put(350.0,850.0){\rule[-0.200pt]{331.237pt}{0.400pt}}
\put(150,700){\makebox(0,0){\Large{$c_{eff}$}}}
\put(1012,25){\makebox(0,0){\large{$l\surd\sigma_k$}}}
\put(350.0,150.0){\rule[-0.200pt]{0.400pt}{168.630pt}}
\put(497.0,509.0){\rule[-0.200pt]{0.400pt}{3.854pt}}
\put(487.0,509.0){\rule[-0.200pt]{4.818pt}{0.400pt}}
\put(487.0,525.0){\rule[-0.200pt]{4.818pt}{0.400pt}}
\put(631.0,496.0){\rule[-0.200pt]{0.400pt}{4.095pt}}
\put(621.0,496.0){\rule[-0.200pt]{4.818pt}{0.400pt}}
\put(621.0,513.0){\rule[-0.200pt]{4.818pt}{0.400pt}}
\put(810.0,472.0){\rule[-0.200pt]{0.400pt}{8.191pt}}
\put(800.0,472.0){\rule[-0.200pt]{4.818pt}{0.400pt}}
\put(800.0,506.0){\rule[-0.200pt]{4.818pt}{0.400pt}}
\put(988.0,525.0){\rule[-0.200pt]{0.400pt}{13.972pt}}
\put(978.0,525.0){\rule[-0.200pt]{4.818pt}{0.400pt}}
\put(978.0,583.0){\rule[-0.200pt]{4.818pt}{0.400pt}}
\put(1256.0,481.0){\rule[-0.200pt]{0.400pt}{16.863pt}}
\put(1246.0,481.0){\rule[-0.200pt]{4.818pt}{0.400pt}}
\put(1246.0,551.0){\rule[-0.200pt]{4.818pt}{0.400pt}}
\put(452.0,517.0){\rule[-0.200pt]{21.681pt}{0.400pt}}
\put(452.0,507.0){\rule[-0.200pt]{0.400pt}{4.818pt}}
\put(542.0,507.0){\rule[-0.200pt]{0.400pt}{4.818pt}}
\put(542.0,505.0){\rule[-0.200pt]{43.121pt}{0.400pt}}
\put(542.0,495.0){\rule[-0.200pt]{0.400pt}{4.818pt}}
\put(721.0,495.0){\rule[-0.200pt]{0.400pt}{4.818pt}}
\put(721.0,489.0){\rule[-0.200pt]{42.880pt}{0.400pt}}
\put(721.0,479.0){\rule[-0.200pt]{0.400pt}{4.818pt}}
\put(899.0,479.0){\rule[-0.200pt]{0.400pt}{4.818pt}}
\put(899.0,554.0){\rule[-0.200pt]{43.121pt}{0.400pt}}
\put(899.0,544.0){\rule[-0.200pt]{0.400pt}{4.818pt}}
\put(1078.0,544.0){\rule[-0.200pt]{0.400pt}{4.818pt}}
\put(1078.0,516.0){\rule[-0.200pt]{86.001pt}{0.400pt}}
\put(1078.0,506.0){\rule[-0.200pt]{0.400pt}{4.818pt}}
\put(497,517){\circle*{24}}
\put(631,505){\circle*{24}}
\put(810,489){\circle*{24}}
\put(988,554){\circle*{24}}
\put(1256,516){\circle*{24}}
\put(1435.0,506.0){\rule[-0.200pt]{0.400pt}{4.818pt}}
\put(613.0,663.0){\rule[-0.200pt]{0.400pt}{12.045pt}}
\put(603.0,663.0){\rule[-0.200pt]{4.818pt}{0.400pt}}
\put(603.0,713.0){\rule[-0.200pt]{4.818pt}{0.400pt}}
\put(778.0,628.0){\rule[-0.200pt]{0.400pt}{8.913pt}}
\put(768.0,628.0){\rule[-0.200pt]{4.818pt}{0.400pt}}
\put(768.0,665.0){\rule[-0.200pt]{4.818pt}{0.400pt}}
\put(999.0,519.0){\rule[-0.200pt]{0.400pt}{15.658pt}}
\put(989.0,519.0){\rule[-0.200pt]{4.818pt}{0.400pt}}
\put(989.0,584.0){\rule[-0.200pt]{4.818pt}{0.400pt}}
\put(1219.0,423.0){\rule[-0.200pt]{0.400pt}{32.521pt}}
\put(1209.0,423.0){\rule[-0.200pt]{4.818pt}{0.400pt}}
\put(1209.0,558.0){\rule[-0.200pt]{4.818pt}{0.400pt}}
\put(558.0,688.0){\rule[-0.200pt]{26.499pt}{0.400pt}}
\put(558.0,678.0){\rule[-0.200pt]{0.400pt}{4.818pt}}
\put(668.0,678.0){\rule[-0.200pt]{0.400pt}{4.818pt}}
\put(668.0,647.0){\rule[-0.200pt]{52.998pt}{0.400pt}}
\put(668.0,637.0){\rule[-0.200pt]{0.400pt}{4.818pt}}
\put(888.0,637.0){\rule[-0.200pt]{0.400pt}{4.818pt}}
\put(889.0,552.0){\rule[-0.200pt]{52.998pt}{0.400pt}}
\put(889.0,542.0){\rule[-0.200pt]{0.400pt}{4.818pt}}
\put(1109.0,542.0){\rule[-0.200pt]{0.400pt}{4.818pt}}
\put(1109.0,491.0){\rule[-0.200pt]{52.998pt}{0.400pt}}
\put(1109.0,481.0){\rule[-0.200pt]{0.400pt}{4.818pt}}
\put(613,688){\circle{24}}
\put(778,647){\circle{24}}
\put(999,552){\circle{24}}
\put(1219,491){\circle{24}}
\put(1329.0,481.0){\rule[-0.200pt]{0.400pt}{4.818pt}}
\put(350,500){\usebox{\plotpoint}}
\put(350.0,500.0){\rule[-0.200pt]{331.237pt}{0.400pt}}
\end{picture}

\end	{center}
\caption{As in Fig.\ref{fig_NGceffN4} but for SU(5) at $\beta=80.0$.}
\label{fig_NGceffN5}
\end 	{figure}
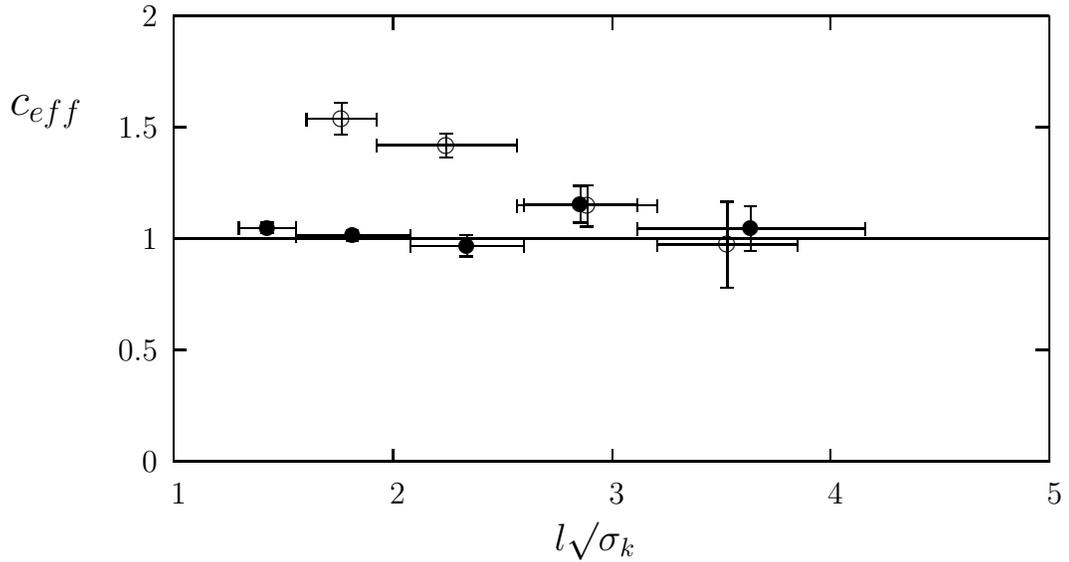

\begin{thebibliography}{99}

\bibitem{gs_k1} 
B. Bringoltz, M. Teper
Phys. Lett. B645 (2007) 383 [hep-th/0611286]. 


\bibitem{es_k1} 
A. Athenodorou, B. Bringoltz, M. Teper
Phys. Lett. B656 (2007) 132 [arXiv:0709.0693].


\bibitem{es_k} 
A. Athenodorou, B. Bringoltz, M. Teper
in preparation.

\bibitem{polyakov}
A. M. Polyakov, {\it Gauge Fields and Strings} (Harwood, 1987). 

\bibitem{thooft}
G. 't Hooft, Nucl. Phys. B72 (1974) 461; B75 (1974) 461.
J. Polchinski, hep-th/9210045.

\bibitem{LSW} 
M. Luscher, K. Symanzik and P. Weisz,
Nucl. Phys. B173 (1980) 365.


\bibitem{PS} 
J. Polchinski and A. Strominger,
Phys. Rev. Lett. 67 (1991) 1681.
 
\bibitem{maldacena} 
J. Maldacena, Adv.Theor. Math. Phys. 2 (1998) 231 [hep-th/9711200].\\
O. Aharony, S. Gubser, J. Maldacena, H. Ooguri and Y. Oz,
Phys. Rept. 323 (2000) 183 [hep-th/9905111].

\bibitem{zamaklar_evans} 
K. Peeters and M. Zamaklar, arXiv:0708.1502. \\
J. Erdmenger, N. Evans, I. Kirsch and E. Threlfall
arXiv:0711.4467. 

\bibitem{arvis}
J. Arvis, Phys. Lett. 127B (1983) 106.

\bibitem{LW} 
M. L\"uscher and P. Weisz, JHEP 0407 (2004) 014
[arXiv:hep-th/0406205].

\bibitem{JD} 
J. Drummond, arXiv:hep-th/0411017; arXiv:hep-th/0608109.\\
N. Hari Dass and P. Matlock,  arXiv:hep-th/0606265,
arXiv:hep-th/0611215.

\bibitem{old_k_oxford} 
B. Lucini, M. Teper 
Phys. Rev. D64 (2001) 105019 [hep-lat/0107007];
Phys. Rev. D66 (2002) 097502 [hep-lat/0206027];
Phys. Lett. B501 (2001) 128 [hep-lat/0012025].

\bibitem{blmtuw-glue04}
B. Lucini, M. Teper, U. Wenger  
JHEP 0406 (2004) 012 [hep-lat/0404008].

\bibitem{old_k_pisa} 
L. Del Debbio, H. Panagopoulos, P. Rossi and E. Vicari,
Phys. Rev. D65 (2002) 021501 [hep-th/0106185];
JHEP 0201 (2002) 009 [hep-th/0111090].

\bibitem{ckahmN}
C. Korthals Altes and H. Meyer, hep-ph/0509018.

\bibitem{nair}   
D. Karabali, C. Kim and V. P. Nair,
Phys. Lett. B434 (1998) 103  [hep-th/9804132];\\
V. P. Nair, Mod. Phys. Lett. A18 (2003) 2415; 
hep-th/0309061.\\
A. Agarwal, D. Karabali, and V. P. Nair,
Phys. Rev. D77 (2008) 025014 [arXiv:0705.2898].

\bibitem{nair_other} 
R. Leigh, D. Minic, A. Yelnikov
Phys. Rev. D76 (2007) 065018 [hep-th/060406]. \\
L. Freidel, R. Leigh, D. Minic, A. Yelnikov,
arXiv:0801.1113.\\ 
R. Leigh and D. Minic, hep-th/0407051.\\
J. Greensite and S. Olejnik,
arXiv:0709.2370; arXiv:0709.2370.\\
P. Orland,  arXiv:0710.3733;
Phys. Rev. D75 (2007) 101702 [arXiv:0704.0940];
Phys. Rev. D71 (2005) 054503 [hep-lat/0501026].

\bibitem{lat07} 
B. Bringoltz and M. Teper, PoS(LATTICE 2007)291
[arXiv:0708.3447].
 
\bibitem{meyerV} 
H. Meyer,  Nucl. Phys. B758 (2006) 204 [hep-lat/0607015]. 

\bibitem{majumdarV} 
N. Hari Dass, P. Majumdar hep-lat/0702019; arXiv:0709.4170;\\
P. Majumdar hep-lat/0406037.

\bibitem{glued3}
M. Teper, Phys. Rev. D59 (1999) 014512 [hep-lat/9804008].


\bibitem{hmmt} 
H. Meyer and M. Teper, JHEP 0412 (2004) 031 [hep-lat/0411039].


\bibitem{gliozzi} 
P. Giudice, F. Gliozzi, S. Lottini PoS (LAT2006) 065
[hep-lat/0609055]; PoS(LATTICE 2007)302 [arXiv:0710.0764].\\
M. Caselle, P. Giudice, F. Gliozzi, P. Grinza, S. Lottini,
arXiv:0710.3522.


\bibitem{CSold} 
J. Ambjorn, P. Olesen and C. Peterson,
Nucl. Phys. B240 (1984) 189, 533; B244 (1984) 262;
Phys. Lett. B142 (1984) 410.

\bibitem{BaliDeldar}
S. Deldar,
Phys. Rev. D62 (2000) 034509 [hep-lat/9911008]; 
JHEP 0101 (2001) 013 [hep-ph/9912428].\\
G. Bali,
Phys. Rev. D62 (2000) 114503 [hep-lat/0006022].

\bibitem{MQCD}
A. Hanany, M. Strassler and A. Zaffaroni,
Nucl. Phys. B513 (1998) 87 [hep-th/9707244]. \\ 
M. Strassler,
Nucl. Phys. Proc. Suppl. 73 (1999) 120 [hep-lat/9810059]. \\
M. Strassler,
Prog. Theor. Phys. Suppl. 131 (1998) 439 [hep-lat/9803009].



\bibitem{mt_largeN02}
M. Teper, Proceedings from the INT vol 12 
(Ed. R. F. Lebed, World Scientific 2002) [hep-ph/0203203].

\bibitem{aamsN}
A. Armoni and M. Shifman, Nucl. Phys. B671 (2003) 67.
[hep-th/0307020];
Nucl. Phys. B664 (2003) 233  [hep-th/0304127].

\bibitem{aabl}
A. Armoni and B. Lucini, JHEP 0606 (2006) 036
[hep-th/0604055].
Nucl.Phys.B664:233-246,2003. 

\end{thebibliography}
\end{document}